\documentclass[structabstract]{aa}  
\usepackage{graphicx}
\usepackage{amssymb,amsmath}
\usepackage{subfig}
\usepackage{graphicx}
\usepackage{epstopdf}
\usepackage{natbib}
\bibpunct{(}{)}{;}{a}{}{,} 
\usepackage{txfonts}
\usepackage{multirow}
\usepackage{hyperref}
\hypersetup{
    pdftoolbar=true,        
    pdfmenubar=true,        
    pdffitwindow=false,     
    pdfstartview={FitH},    
    pdftitle={The strongest gravitational lenses: II. Is the large Einstein radius of MACS J0717.5+3745 in conflict with LCDM?},    
    pdfauthor={Jean-Claude Waizmann},     
    pdfsubject={Draft},   
    pdfcreator={Jean-Claude Waizmann},   
    pdfproducer={Jean-Claude Waizmann}, 
    colorlinks=true,       
    linkcolor=blue,          
    citecolor=blue,        
    filecolor=blue,      
    urlcolor=blue           
}

\graphicspath{{./figures/}}

\newcommand{\dd}{\mathrm{d}}
\newcommand{\e}{\mathrm{e}}

\begin{document}

\title{The strongest gravitational lenses: II. Is the large Einstein 
radius of MACS J0717.5+3745 in conflict with $\Lambda$CDM?}

\titlerunning{The strongest gravitational lenses II.}


\author{J.-C. Waizmann \inst{1,2,3}, M. Redlich \inst{4,5} \and 
M. Bartelmann \inst{4}}

\institute{Dipartimento di Astronomia, Universit\`{a} di Bologna, 
via Ranzani 1, 40127 Bologna, Italy\\ 
\email{jcwaizmann@oabo.inaf.it}
\and
INAF - Osservatorio Astronomico di Bologna, via Ranzani 1, 40127 
Bologna, Italy
\and
INFN, Sezione di Bologna, viale Berti Pichat 6/2, 40127 Bologna, Italy
\and
Zentrum f\"ur Astronomie der Universit\"at Heidelberg, Institut f\"ur 
Theoretische Astrophysik, Albert-Ueberle-Str.~2, 69120 Heidelberg, Germany
\and
The Sydney Institute for Astronomy, The University of Sydney, School of 
Physics A28, NSW 2006, Australia (present address)
}

\authorrunning{J.-C. Waizmann et al.}

\date{\emph{Received 3 July 2012~\slash ~Accepted 7 August 2012}}

\abstract
{With the amount and quality of galaxy cluster data increasing, the question arises 
whether or not the standard cosmological model can be questioned on the basis 
of a single observed extreme galaxy cluster. Usually, the word \textit{extreme} 
refers directly to cluster mass, which is not a direct observable 
and thus subject to substantial uncertainty. Hence, it is desirable 
to extend studies of extreme clusters to direct observables, such as the Einstein radius.}
{We aim to evaluate the occurrence probability of the large observed Einstein 
radius of MACS J0717.5+3745 within the standard $\Lambda$CDM cosmology. In particular, 
we want to model the distribution function of the single largest Einstein radius in 
a given cosmological volume and to study which underlying assumptions and effects 
have the strongest impact on the results.}
{We obtain this distribution by a Monte Carlo approach, based on the semi-analytic modelling of the halo population 
on the past lightcone. After sampling the distribution, we fit the results with the 
general extreme value (GEV) distribution which we use for the subsequent analysis.}
{We find that the distribution of the maximum Einstein radius is particularly sensitive 
to the precise choice of the halo mass function, lens triaxiality, the inner slope of the 
halo density profile and the mass-concentration relation. Using the distributions so obtained,
we study the occurrence probability of the large Einstein radius of MACS J0717.5+3745,
finding that this system is not in tension with $\Lambda$CDM. We also 
find that the GEV distribution can be used to fit very accurately the sampled distributions 
and that all of them can be described by a (type-II) Fr\'{e}chet distribution.}
{With a multitude of effects that strongly influence the distribution of the single largest 
Einstein radius, it is more than doubtful that the standard $\Lambda$CDM cosmology 
can be ruled out on the basis of a single observation. If, despite the large uncertainties 
in the underlying assumptions, one wanted to do so, a much larger Einstein radius 
($\ga 100\arcsec$) than that of MACS J0717.5+3745 would have to be observed.}

\keywords{gravitational lensing: strong -- methods: statistical -- galaxies: clusters:
 general -- galaxies: clusters:individual: MACS J0717.5+3745  -- cosmology: 
 miscellaneous}

\maketitle

\section{Introduction}\label{sec:intro}
Galaxy clusters are extreme objects from many points of view. They are the most 
massive gravitationally bound systems in the Universe and, hence, flag the rarest 
peaks of the initial density field. The gas contained in their gravitational potential 
wells is heated up to extremely high temperatures of the order of $10^7-10^8\,K$, 
resulting in the emission of X-ray radiation. Furthermore, they can give rise to 
spectacular events of strong gravitational lensing. Individually and as a population,
galaxy clusters contain rich information on the formation of structure 
in the Universe that will be recovered to greater extent in the near future by ongoing 
and upcoming surveys, like SPT \citep{Carlstrom2011}, \textit{eROSITA} 
\citep{Cappelluti2011} and \textit{EUCLID} \citep{Laureijs2011}.

Recently, the interest in the extremest among the extreme, the most massive clusters, 
has substantially increased. This development was mainly triggered by the detection 
of very massive galaxy clusters at high redshifts, like XMMU J2235.3$-$2557 at $z = 1.4$ 
\citep{Mullis2005,Rosati2009, Jee2009}, ACT-CL J0102 at $z=0.87$ 
\citep{Marriage2011, Menanteau2011} and SPT-CL J2106 at $z=1.132$ \citep{Foley2011, 
Williamson2011}. Several works studied the probability to find such objects in a standard 
$\Lambda$CDM cosmology \citep{Holz2010, Baldi2011, Cayon2011, Hotchkiss2011, 
Mortonson2011, Chongchitnan2012, Waizmann2012a}. All these studies 
focused on the mass of galaxy clusters, which is unfortunately not a direct observable. 
The mass of a galaxy cluster, ill defined in the first place, is subject to substantial scatter 
and biases. Hence, it is desirable to study extremes in direct, better defined, observables, 
such as strong lensing signals. 

A particular interesting case from this point of view is the extremely large critical curve 
of the X-ray luminous galaxy cluster MACS~J0717.5+3745 at redshift $z=0.546$, which 
has been independently detected by the Massive Cluster Survey (MACS) \citep{Ebeling2001, 
Ebeling2007} and as a host of a diffuse radio source \citep{Edge2003}. A strong-lensing 
analysis revealed that the effective Einstein radius, with $\theta_{\rm eff}=(55\pm3)\arcsec$ 
for an estimated source redshift of $z\simeq2.5$, is the largest known at redshifts of 
$z>0.5$ \citep{Zitrin2009, Zitrin2011a}. It is unclear whether or not such a large Einstein 
radius is consistent with the $\Lambda$CDM cosmology \citep{Zitrin2009}.  

In this work, we study the distributions of the single largest Einstein radius by means of 
semi-analytic modelling of the halo distribution on the past lightcone, as introduced 
in \cite{Redlich2012}. On the basis of the study of \cite{Oguri2003}, the haloes are modelled 
using triaxial density profiles. By Monte Carlo (MC) sampling the distribution of the maximum 
for different underlying assumptions like the mass function, the allowed range of triaxiality, 
the inner slope and the mass-concentration relation, we study the impact of different choices 
on the resulting distributions of the maximum. We use the results to assess the occurrence 
probability of the Einstein radius of MACS~J0717.5+3745 in the redshift range of 
$0.5\le z\le 1.0$ and fit the generalized extreme value (GEV) distribution to the sampled 
distribution of the maximum Einstein radius.   

This paper is structured as follows. In \autoref{sec:SL_intro}, we 
briefly review the basics of strong cluster lensing followed by an introduction of the 
semi-analytic modelling of the distribution of Einstein radius in \autoref{sec:samER}. 
In \autoref{sec:evsER}, we introduce extreme values statistics as far as it is relevant for the 
presented work before studying in detail the distribution of the largest Einstein radius 
for different underlying physical assumptions in \autoref{sec:distriER}. Afterwards, we 
perform a case study for MACS~J0717.5+3745 in \autoref{sec:MACSJ0717} and close 
with a summary and the conclusions in \autoref{sec:conclusions}.

Throughout this work we adopt the \textit{Wilkinson Microwave Anisotropy Probe 7--year} 
(WMAP7) parameters $(\Omega_{{\rm m}0}, \Omega_{\Lambda 0}, \Omega_{{\rm b}0}, h, 
\sigma_8) = (0.727, 0.273, 0.0455, 0.704, 0.811)$ \citep{Komatsu2011}.
\section{Strong lensing by galaxy clusters}\label{sec:SL_intro}
The theory of gravitational lensing is well established and has become a very important 
tool for the study of the dark components of the Universe \citep[for recent reviews, see 
e.g.][]{Bartelmann2010, Kneib2011}. In this work, we focus on strong lensing by 
galaxy clusters and in particular on the distribution of Einstein radii.

Einstein radii measure the size of the tangential critical curve, defined as the curve of 
vanishing tangential eigenvalue
\begin{equation}
\lambda_t(\vec{\theta_{\rm t}})=1-\kappa -\gamma\stackrel{!}{=}0 \;,
\label{eq:tc_curve}
\end{equation}
with $\kappa$ and $\gamma$ denoting the convergence and the shear, respectively.
In the axially symmetric case, the shear can be written as
\begin{equation}
	\gamma(\theta) = \overline\kappa(\theta)-\kappa(\theta) \;,
	\label{eq:gammacirc}
\end{equation}
where $\overline\kappa(\theta)$ denotes the mean convergence within a circle of radius 
$\theta$. One can then define the Einstein radius as the radius of a circle enclosing a 
mean convergence of unity 
\begin{equation}
	1-\overline\kappa(\theta_{\rm E})=0 \;,
	\label{eq:thetae}
\end{equation}
where $\theta_{\rm E}$ denotes the Einstein radius. Recalling the relation between the 
convergence and the surface mass density $\kappa=\Sigma / \Sigma_{\rm crit}$, one 
can formulate the definition of $\theta_{\rm E}$ as 
\begin{equation}
	\overline\kappa(\theta_{\rm E})=\frac{\overline\Sigma(\theta_{\rm E})}
	{\Sigma_{\rm crit}}=1 \;,
\end{equation}
stating that the mean surface density $\overline\Sigma(\theta_{\rm E})$ within the 
Einstein radius equals the critical surface mass density $\Sigma_{\rm crit}$.

The assumption of axially symmetric lenses is unsustainable for realistic lenses, and 
thus several definitions of the Einstein radius for the case of arbitrary lenses with irregular 
tangential critical curves exist.

In this work, we focus on two definitions: the first, introduced by \cite{Meneghetti2011} 
and known as the median Einstein radius, is defined as
\begin{equation}
\theta_{\rm median}\equiv\text{median}\left\lbrace\sqrt{\left(\theta_{i,x}-
\theta_{c,x}\right)^2+\left(\theta_{i,y}-\theta_{c,y}\right)^2} \,\mid\,\vec{\theta_i}\in 
\vec{\theta_{\rm t}}\right\rbrace\;,
\end{equation}
where $\vec{\theta_{\rm t}}$ denotes the set of the tangential critical points and 
$\vec{\theta_{\rm c}}$ is the centre of the lens. The second definition is of geometrical 
nature and is usually referred to as effective Einstein radius. It is defined as
\begin{equation}
\theta_{\rm eff}\equiv\sqrt{\frac{A}{\pi}}\;,
\end{equation}
where $A$ is the area enclosed by the critical curve. Of course, both definitions 
are identical to the original definition of $\theta_{\rm E}$ in the axially symmetric case. 
In the next section, we will discuss in detail how the distributions of these two quantities 
can be modelled in a cosmological context.
\section{Semi-analytic modelling of the distribution of Einstein radii}\label{sec:samER}
For the semi-analytic modelling of the distribution of Einstein radii, we closely follow 
the work of \cite{Redlich2012}, which provides a more detailed 
discussion of the algorithm, including the computation of the lensing signal and the 
cosmological population of haloes. Therefore, we will in this section only briefly 
summarise those aspects of the semi-analytic modelling that will be needed to follow the 
remainder of this work. However, in our paper, we conservatively neglect the impact of 
mergers on the extreme value distribution. The results of \cite{Redlich2012} indicate that 
the inclusion of mergers can be expected to significantly shift the distribution to larger 
Einstein radii. Thus, the results presented in our paper can be considered as conservative 
estimates.
\begin{figure*}
\centering
\includegraphics[width=0.495\linewidth]{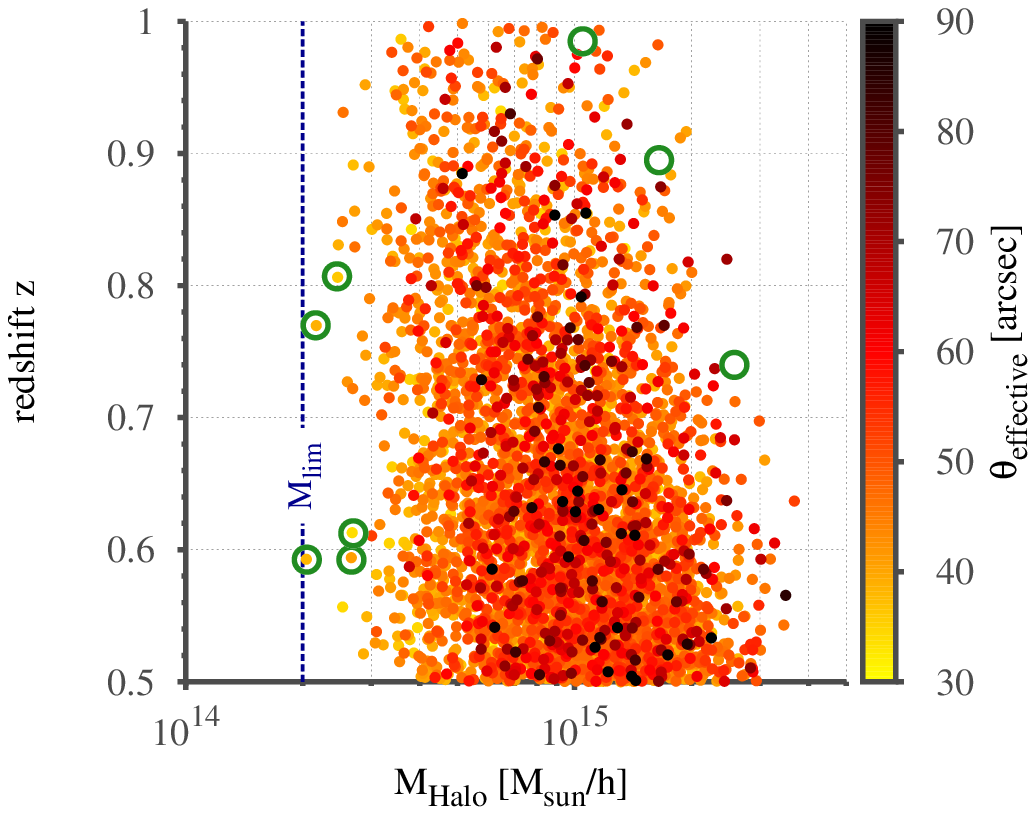} 
\includegraphics[width=0.495\linewidth]{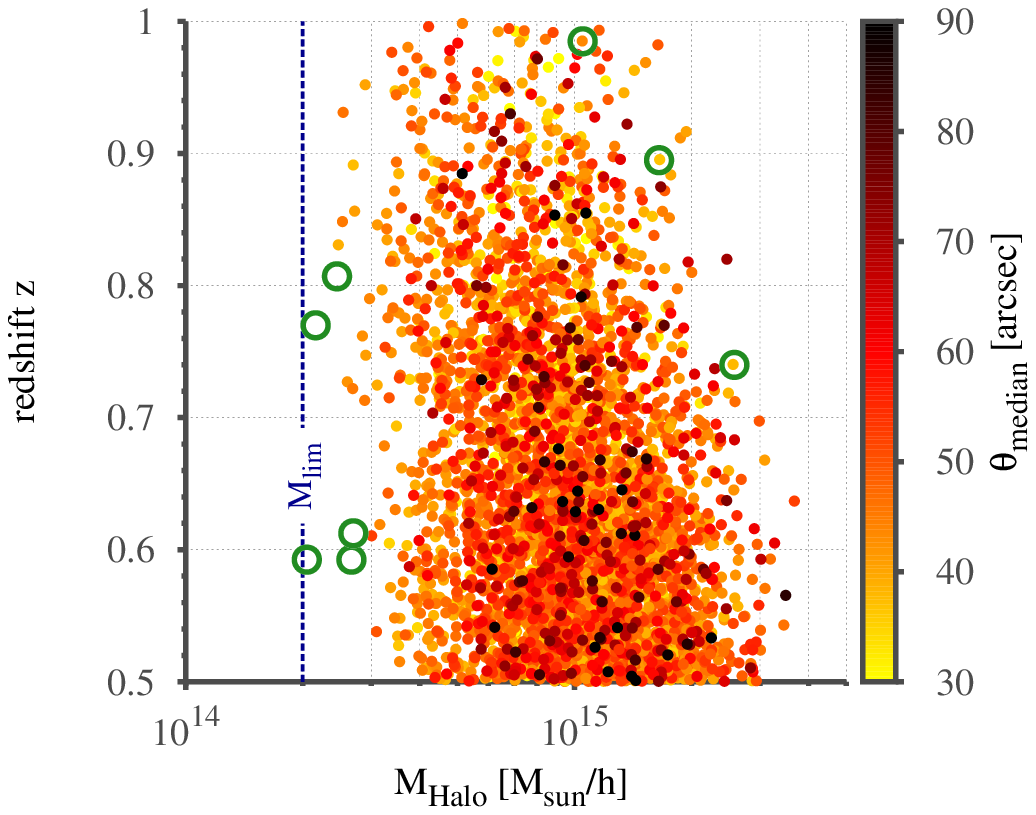}
\caption{Distribution in mass and redshift of $4\,000$ maxima of the effective (left panel) 
and the median Einstein radius (right panel) in the redshift interval of $0.5\le z\le 1.0 $ on 
the full sky based on the \cite{Tinker2008} mass function.The colour encodes the size 
of the individual largest Einstein radius from each simulation run. The green circles denote 
exemplary systems for which the two Einstein radius definitions selected different haloes 
as the maximum.}
\label{fig:Mz_distribution}
\end{figure*}
\subsection{Modelling triaxial haloes}\label{sec:singleHaloes}
The integral part for the modelling of the Einstein radius distribution is the inclusion of 
triaxiality as discussed, for instance, in \citet[][hereafter OB09]{Oguri2009}, which is based on 
the work of \citet[][hereafter JS02]{Jing2002}. In their work, JS02 generalised the 
\citet[][hereafter NFW]{nfw1996} profile to a  triaxial model, where the axis ratios for a given 
mass, $M$, and redshift, $z$, can be sampled from the following empirically derived probability 
density functions, assuming the ordering $\left( a \leq b \leq c\right)$ 
\begin{align}
\label{eq:p_a}
p(a/c) &= \frac{1}{\sqrt{2\pi}\sigma_s} \exp\left[-\frac{(a_{\mathrm{sc}}-0.54)^2}
{2\sigma_{\mathrm{s}}^2}\right]\frac{\mathrm{d}a_{\mathrm{sc}}}{\mathrm{d}(a/c)} \; , \\ 
p(a/b|a/c) &= \dfrac{3}{2(1-r_{\mathrm{min}})}\left[1-\left(\dfrac{2a/b-1-r_{\mathrm{min}}}
{1-r_{\mathrm{min}}}\right)^2\right]  \; ,\label{eq:p_ab}
\end{align}
where the latter relation holds for $a/b \ge r_{\mathrm{min}}$ and is zero otherwise and
\begin{equation}\label{eq:a_sc}
a_{\mathrm{sc}} = \frac{a}{c} \left( \frac{M}{M_*} \right)^{0.07[\Omega_{{\rm m}}(z)]^{0.7}} \; , 
\quad r_{\mathrm{min}} = \operatorname{max}\left(a/c,0.5\right) \; . 
\end{equation}
Here, $M_*$ is the characteristic non-linear mass scale and, according to JS02, the best-fitting 
parameter for the width of the axis-ratio distribution $p(a/c)$ is $\sigma_{\rm s} = 0.113$.

The concentration parameter $c_{\rm e}$ is defined as $c_{\rm e}\equiv R_{\rm e}/R_0$,
where $R_{\rm e}$ is determined such that the mean density within the
ellipsoid of the major axis radius $R_{\rm e}$ is $\Delta_{\rm e}\Omega(z)\rho_{\rm crit}(z)$,
with
\begin{equation}
5\Delta_{\rm vir}(z)\left(c^2/ab\right)^{0.75},
\end{equation}
where $\Delta_{\rm vir}(z)$ is the overdensity of objects virialized at redshift $z$, which we 
approximate according to \cite{Nakamura1997}. In their work, JS02 found a log-normal 
distribution for the concentration,
\begin{equation}
 p(c_{\rm e})=\frac{1}{\sqrt{2\pi}\sigma_{\rm c}} \exp\left[
-\frac{(\ln c_{\rm e}-\ln \bar{c}_{\rm e})^2}{2\sigma_{\rm c}^2}\right]\frac{1}{c_{\rm e}}, 
\label{eq:p_ce}
\end{equation}
with a dispersion of $\sigma_{\rm c} = 0.3$. Following \cite{Oguri2003}, we include a 
correlation between the axis ratio $a/c$ and the mean concentration,
\begin{align}
\label{eq:ce}
\bar{c}_{\mathrm{e}} &= f_{\rm c} A_{\rm e} \sqrt{\frac{\Delta_{\mathrm{vir}}(z_{\rm c})}
{\Delta_{\mathrm{vir}}(z)}} \left( \frac{1 + z_{\rm c}}{1 + z} \right) \; , \\
f_{\rm c} &= \mathrm{max} \left\{0.3, 1.35 \exp \left[ - \left(\frac{0.3}{a_{\mathrm{sc}}} 
\right)^2 \right] \right\} \; ,
\label{eq:fc}
\end{align}
where $z_{\rm c}$ is the collapse redshift. In order to avoid unrealistically small 
concentrations for particularly small axis ratios $a_{\rm sc}$, we use the correction 
introduced by OB09, forcing $f_{\rm c} \ge 0.3$ in \autoref{eq:fc}. Obviously, triaxial 
haloes with particularly small axis ratios $a_{\mathrm{sc}}$ (and hence also small 
concentrations $c_{\rm e}$) are highly elongated objects whose lens potential is 
dominated by masses well outside the virial radius \citep[see e.g.][]{Oguri&Keeton2004}. 
There are two ways of dealing with the problem of these unrealistic scenarios. The first 
is based on truncating the density profile beyond the virial radius (see e.g. 
\cite{Baltz2009}; OB09), the second approach suppresses particularly small axis ratios, 
$a_{\mathrm{sc}}$, from the tail of the underlying axis ratio distribution (see 
\autoref{sub:triaxiality}). Following JS02, we set the free parameter $A_{\rm e} = 1.1$ for 
a standard $\Lambda$CDM model, unless stated otherwise. The expressions listed so far 
are valid for an inner slope of the density profile of $\alpha_{\rm NFW} = 1.0$. For the 
case of $\alpha_{\rm NFW} = 1.5$, we use the simple relation $\bar{c}_{\rm e}
(\alpha_{\rm NFW} = 1.5) = 0.5 \times \bar{c}_{\rm e}(\alpha_{\rm NFW} = 1.0)$ 
(\citet{Keeton&Madau2001}; JS02).
\subsection{Preparatory considerations for the MC sampling}\label{sec:prepConsid}
In order to obtain the extreme value distribution of the maxima of the Einstein radii, 
we have to sample the cluster populations for many mock realisations and to collect 
the largest Einstein radius from each realisation. In view of this, the choice of the 
redshift interval and the allowed halo mass range is decisive to keep computational 
costs under control. 

On the basis of the strong-lensing analysis of $12$ MACS clusters in the interval 
$0.5\le z\le 0.6$ by \cite{Zitrin2011a}, we focus our study of the distribution of 
the largest Einstein radii on clusters in the redshift range of $0.5\le z \le 1$, assuming 
a source redshift of $z_{\rm s}=2.0$. This choice already drastically reduces the 
number of haloes that have to be simulated.
The remaining task is to identify a lower mass limit $M_{\rm lim}$, such that the 
inferred sampled maxima distribution is not biased. To do so, we simulated $4\,000$ 
maxima with $M>2\times 10^{14}\,M_\odot/h$ in $0.5\le z \le 1$ and present the 
results in \autoref{fig:Mz_distribution} for both the effective and the median Einstein 
radii. It can be directly inferred from the distribution of the maxima that 
$M_{\rm lim}=2\times 10^{14}\,M_\odot/h$ is a sufficient lower mass limit, 
confirming the results of OB09. Thus, we will adopt this value throughout 
this work, unless stated otherwise.

The distribution of the maxima in mass and redshift, presented in 
\autoref{fig:Mz_distribution}, shows that the maxima stem from a wide range of masses. 
It is not unlikely that a rather low mass cluster gives rise to the largest 
Einstein radius. The fact that most of the maxima are found in the lower redshift range 
is a consequence of the selected lensing geometry determined by the choice of the source 
redshift $z_{\rm s}=2.0$. Since we are modelling triaxial haloes, this is a first indication that 
the orientation of the halo with respect to the observer, the lensing geometry and the 
concentration are more important than the mass. The green circles in 
\autoref{fig:Mz_distribution} denote systems for which the Einstein radius definitions 
select different haloes to exhibit the maximum Einstein radius. We note that the extremely 
large Einstein radii (black dots) are not affected by the choice of the Einstein radius 
definition.

\begin{figure*}
\centering
\includegraphics[width=0.425\linewidth]{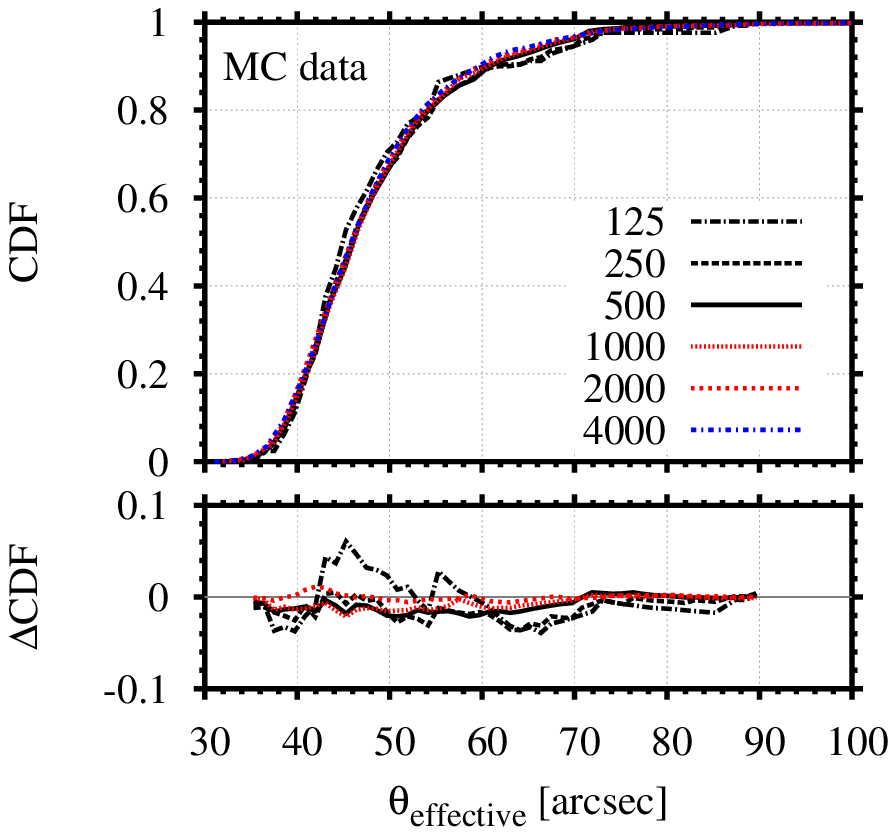}
\includegraphics[width=0.425\linewidth]{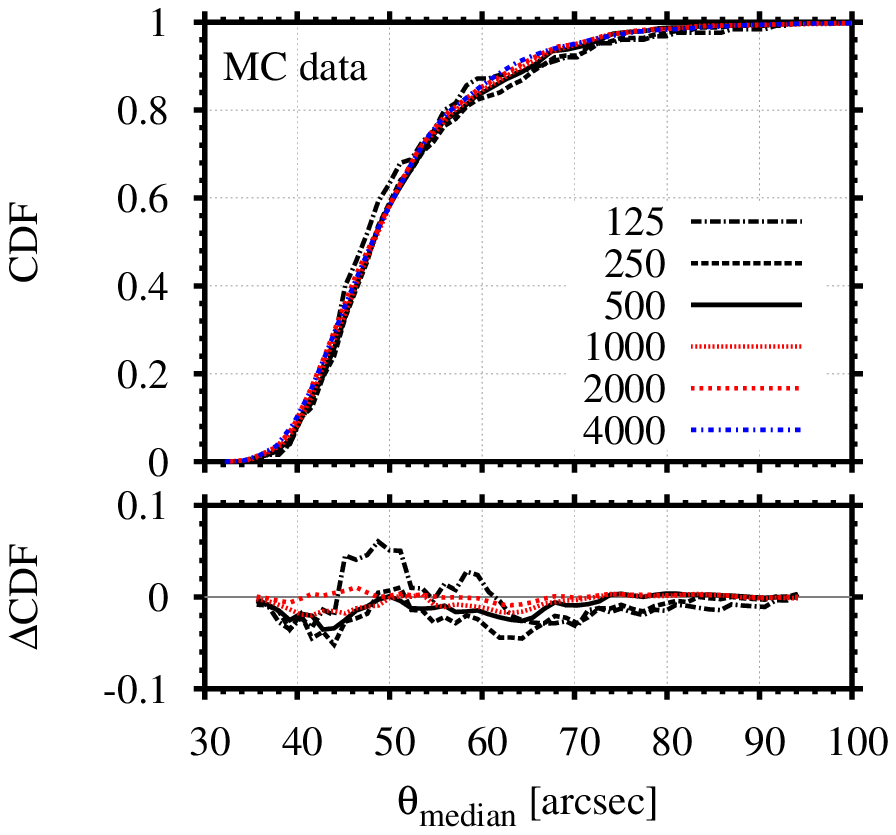}\\
\includegraphics[width=0.425\linewidth]{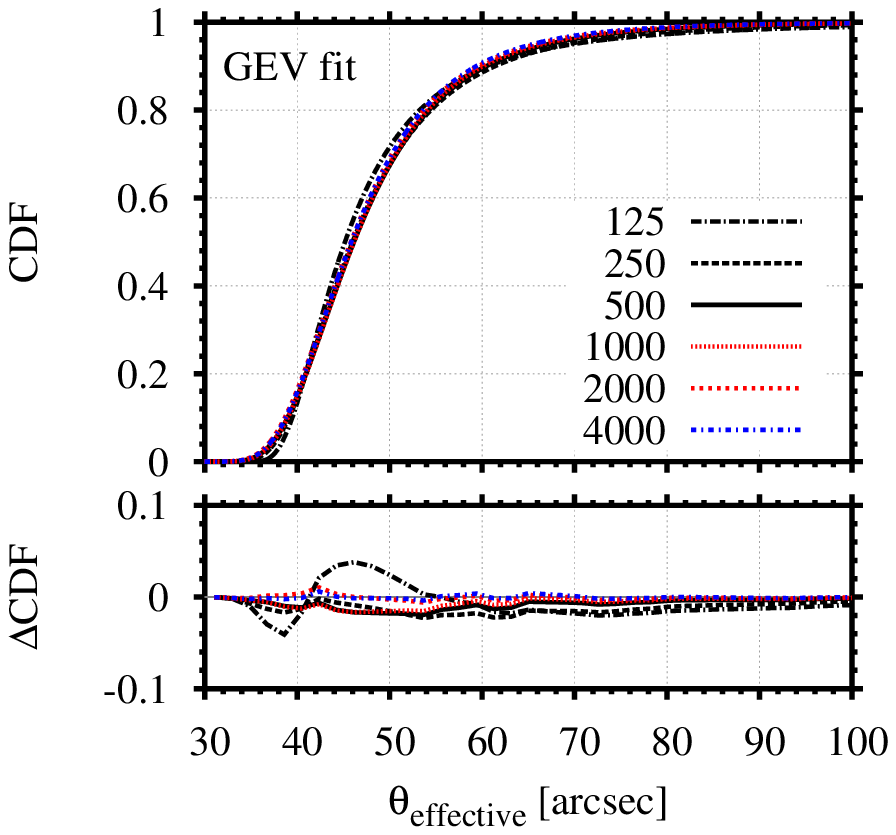}
\includegraphics[width=0.425\linewidth]{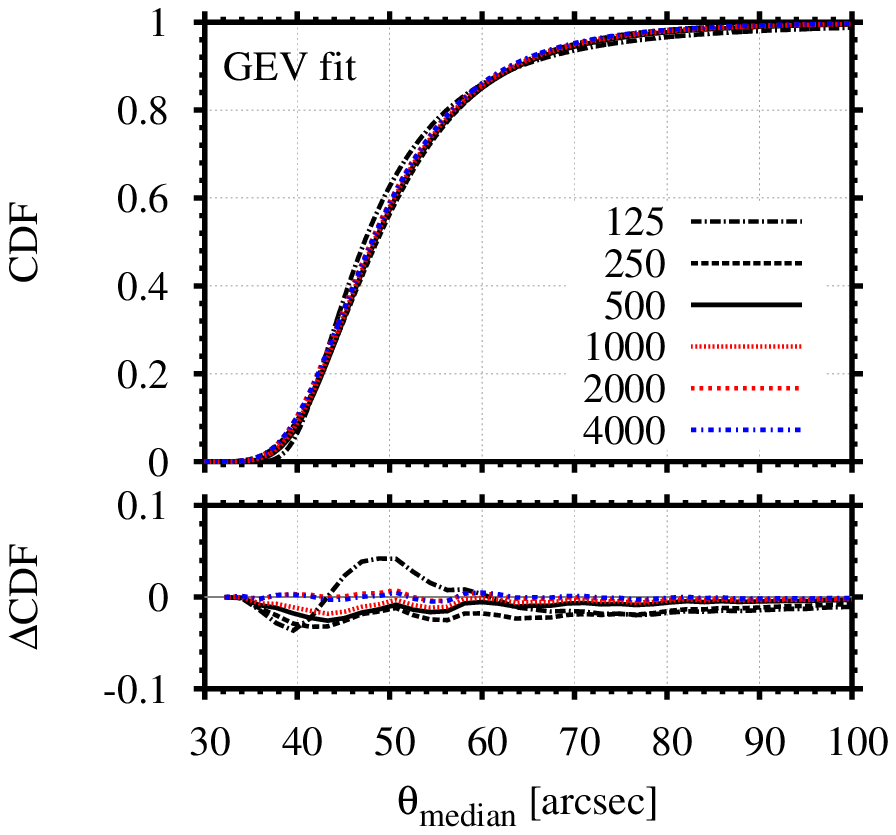}
\caption{CDFs of the largest effective (left panels) and median (right panels) Einstein radius 
for a different number of maxima in the range between $125$ and $4\,000$, 
assuming the \cite{Tinker2008} mass function, $M_{\rm lim}=2\times 10^{14}\,M_\odot /h$ 
and the redshift interval of $0.5\le z\le 1.0 $ on the full sky. The upper row shows 
the CDFs directly based on the MC simulations, and the lower row presents the 
corresponding fits of the GEV distribution. The small lower panels show the 
difference, $\Delta$CDF, with respect to the high-resolution run based on $4\,000$ 
maxima.}\label{fig:nRun}
\end{figure*}

Having fixed the mass and redshift range, the last step in optimizing the 
computational cost is to understand how many maxima actually have to be sampled 
in order to construct the cumulative distribution function (CDF) of the largest effective 
and median Einstein radii. To this end, we computed the respective CDFs for 
different sample sizes $N_{\rm samp}$ between $125$ and $4\,000$. Each CDF was 
computed at $50$ linearly equidistant points between the largest and 
the smallest value. The results of these computations are presented in the upper 
row of \autoref{fig:nRun}, where the CDFs themselves are shown in the upper 
panels and the differences of the CDFs with respect to the highest 
resolution run ($N_{\rm samp}=4\,000$) are shown in the small lower panels. As expected, the 
noise of the CDFs decreases with increasing $N_{\rm samp}$. For $N_{\rm samp}\geq1\,000$, the 
difference with respect to the high resolution run is $\lesssim 0.02$, corresponding to 
an over-estimation of the occurrence probability of a given Einstein radius by less than 
two per cent. Hence, we will utilise $N_{\rm samp}=1\,000$ for all of our computations 
in the remainder of this work, unless stated otherwise. 

\begin{table*}
\centering
\caption{Values of the location, scale and shape parameters, $\alpha_{\rm eff}$, $\beta_{\rm eff}$ and 
$\gamma_{\rm eff}$,  of the fitted GEV distributions of the maximum effective Einstein radius for 
different sample sizes $N_{\rm samp}$,  as shown in the lower left panel of \autoref{fig:nRun}. 
In addition, the number of degrees of freedom (DoF) of the fit and the root mean square (rms) of the 
residuals are given.}
\begin{tabular}{rccccc} \hline\hline
$N_{\rm samp}$ & $\alpha_{\rm eff}$ & $\beta_{\rm eff}$ & $\gamma_{\rm eff}$ & 
DoF\tablefootmark{a} & RMS of Residuals  \\
\hline
$125$ & $43.17\pm 0.066$ & $5.12\pm 0.101$ & $0.35\pm 0.020$ & $47$
 & $1.234\times 10^{-2}$ \\ 
$250$ & $43.69\pm 0.050$ & $6.04\pm 0.077$ & $0.23\pm 0.014$ & $47$
 & $8.665\times 10^{-3}$ \\
$500$ & $43.81\pm 0.044$ & $6.19\pm 0.067$ & $0.15\pm 0.012$ & $41$
 & $6.394\times 10^{-3}$  \\
$1000$ & $43.83\pm 0.026$ & $6.16\pm 0.039$ & $0.14\pm 0.007$ & $41$
 & $3.736\times 10^{-3}$  \\
$2000$ & $43.52\pm 0.017$ & $6.14\pm 0.026$ & $0.13\pm 0.005$ & $41$
 & $2.442\times 10^{-3}$  \\
$4000$ & $43.58\pm 0.015$ & $6.07\pm 0.023$ & $0.13\pm 0.004$ & $34$
 & $2.036\times 10^{-3}$ \\
\hline
\end{tabular}\label{tab:nrun_param}
\tablefoot{ \tablefoottext{a} The sampled CDF is calculated at 50 equally linearly spaced points 
between the largest and the smallest value. }
\end{table*}
\section{Applying extreme value statistics to the distribution of the largest Einstein radii}\label{sec:evsER}
Extreme value statistics (EVS) (for an introduction, see e.g. \cite{Gumbel1958, 
Kotz2000, Coles2001, Reiss2007}) models the stochastic behaviour of the 
extremes and tries to give a quantitative answer to the question of how frequent 
unusual observations are. In the framework of EVS, there are two approaches to 
the modelling of rare events. The first one, also known as the Gnedenko approach 
\citep{Fisher1928, Gnedenko1943}, models the distribution of the block maxima, 
while the second one, known as the Pareto approach \citep{Pickands1975}, models 
the distribution of excesses over high thresholds. Since we are interested in the 
study of the distribution of the largest Einstein radii, we will discuss the first 
approach in more detail in the following.
\subsection{The Gnedenko approach}\label{sec:Gnedenko}
This approach is concerned with the modelling of the block maxima $ M_n $
of independently identically distributed (i.i.d.) random variables $X_i$, which are 
defined as
\begin{equation}
M_n=\max(X_1,\dotsc X_n).
\label{eq:blockmax}
\end{equation}
It has been shown \citep{Fisher1928, Gnedenko1943} that, for $n\rightarrow\infty$, 
the limiting CDF of the renormalised block maxima is given by one of the extreme 
value families: Gumbel (Type I), Fr\'{e}chet (Type II) or Weibull (Type III). These 
three families can be unified \citep{vonmises1954, Jenkinson1955} as a general extreme value (GEV) distribution 
\begin{equation}\label{eq:GEV}
  G_{\gamma,\,\beta,\,\alpha}(x) = \left\{ 
  \begin{array}{l l}
    \exp{\left\lbrace -\left[1+\gamma \left(\frac{x-\alpha}{\beta}\right)\right]^
    {-1/\gamma}\right\rbrace}, & \quad {\rm for}\quad\gamma\neq 0,\\
    \exp{\left\lbrace \e^{-\left(\frac{x-\alpha}{\beta}\right)}\right\rbrace} ,& \quad 
    {\rm for}\quad\gamma = 0,\\
  \end{array} \right.
\end{equation}
with the shape, scale and location parameters $\gamma$, $\beta$ and $\alpha$. 
In this generalisation, $\gamma=0$ corresponds to the Type I,  $\gamma>0$ to 
Type II and $\gamma<0$ to the Type III distributions. The corresponding probability 
density function (PDF) is given by $g_{\gamma,\,\beta,\,\alpha}(x)=\dd G_{\gamma,
\,\beta,\,\alpha}(x)/ \dd x$ and reads for the case of $\gamma\neq 0$,
\begin{equation}\label{eq:GEV_pdf} 
\begin{split} 
g_{\gamma,\,\beta,\,\alpha}(x)& = \frac{1}{\beta}\left[1+\gamma 
\left(\frac{x-\alpha}{\beta}\right)\right]^{-1-1/\gamma} \\
& \quad \times\exp{\left\lbrace -\left[1+\gamma 
\left(\frac{x-\alpha}{\beta}\right)\right]^{-1/\gamma}\right\rbrace}.
\end{split} 
\end{equation}
From now on we will adopt the convention that capital initial letters denote the CDF 
(like $G_{\gamma,\,\beta,\,\alpha}(x)$) and small initial letters denote the PDF (like 
$g_{\gamma,\,\beta,\,\alpha}(x)$). The mode, the most likely value, of the GEV 
distribution reads
\begin{equation}
 x_0=\alpha+\frac{\beta}{\gamma}\left[\left(1+\gamma\right)^{-\gamma}-1\right],
\end{equation}
and the expected value is given by
\begin{equation}
\mathrm{E}_{\rm GEV}=\alpha-\frac{\beta}{\gamma}+\frac{\beta}{\gamma}\Gamma
\left(1-\gamma\right),
\end{equation}
 where $\Gamma$ denotes the Gamma function.
\subsection{GEV and the distribution of Einstein radii}\label{sec:GEVandER}
The MC approach to the distribution of the single largest Einstein radius, 
introduced in \autoref{sec:samER}, provides simulated CDFs that are 
discrete by nature. Due to the complexity of the modelling of the Einstein radius 
distribution, it is not possible to find analytic relations for the GEV parameters, as 
can be done for halo masses \citep{Davis2011, Waizmann2011}. Hence, we use 
the limiting GEV distribution from \autoref{eq:GEV} to fit the sampled distributions in 
order obtain analytic relations for the distribution of the largest Einstein radii. 

The GEV distribution, given by \autoref{eq:GEV}, fits the MC-simulated 
distributions very well, as can be inferred from the lower row of \autoref{fig:nRun}. 
In the small panels below, we show the difference $\Delta$CDF of the CDFs 
with respect to the fits based on the run with $N_{\rm samp}=4\,000$. It can be 
seen that the fitted functions deviate less from the high-resolution run with 
respect to the MC data. The blue, dashed-dotted line depicts the deviation of the 
fit from the corresponding data set for $N_{\rm samp}=4\,000$, showing that 
the GEV-based fits are capable of describing the MC simulated distributions of 
the Einstein radii very well. We also present in \autoref{tab:nrun_param} all fitted 
GEV parameters as well as the root mean square of the residuals for the case of the 
effective Einstein radius.  In what follows, we will use the GEV fits for any subsequent 
analysis, like the calculation of PDFs, modes or quantiles.\\
The fitted shape parameters $\gamma$ for all distributions discussed in this 
work are found to be in the range of $0.05<\gamma < 0.2$, which means that 
the distribution of the largest Einstein radii can in general be described by a 
Fr\'{e}chet (Type II) distribution, indicating that the distribution is bounded from 
below. An exception to this will be discussed further in \autoref{sub:triaxiality}. 
The location parameter $\alpha$ is always very close to the mode, the most likely 
maximum, with the two values differing only by roughly one per cent. It is noteworthy 
that the location parameter $\alpha$ can be estimated very well with rather small 
sample sizes, whereas the shape parameter $\gamma$ is subject to larger 
uncertainties. Even for only $125$ samples, the difference of the mode 
with respect to the $N_{\rm samp}=4\,000$ case is less than $2\,{\rm arcsec}$. 
This result is similar to the findings of \cite{Waizmann2011}, who report 
the same behaviour for the case of halo masses.
\section{The distribution of the largest Einstein radii}\label{sec:distriER}
In this section, we study the impact of several underlying assumptions, like mass 
function, triaxiality, inner slope and mass--concentration relation, on the distribution 
of the single largest Einstein radius. 
\subsection{The impact of the mass function}\label{sub:massfunction}
The choice of the mass function must have an effect on the distribution of the largest 
Einstein radii, since it alters the size of the halo population from which the maxima are 
drawn. This effect is particularly important for galaxy clusters since the exponentially 
suppressed tail of the mass function is naturally very sensitive to modification. As shown in 
\autoref{fig:Mz_distribution}, the maxima stem from a relatively broad range of masses. 
Hence, the larger the halo population in this mass range, the more likely it is to sample 
a particularly large Einstein radius.

\begin{figure*}
\centering
\includegraphics[width=0.425\linewidth]{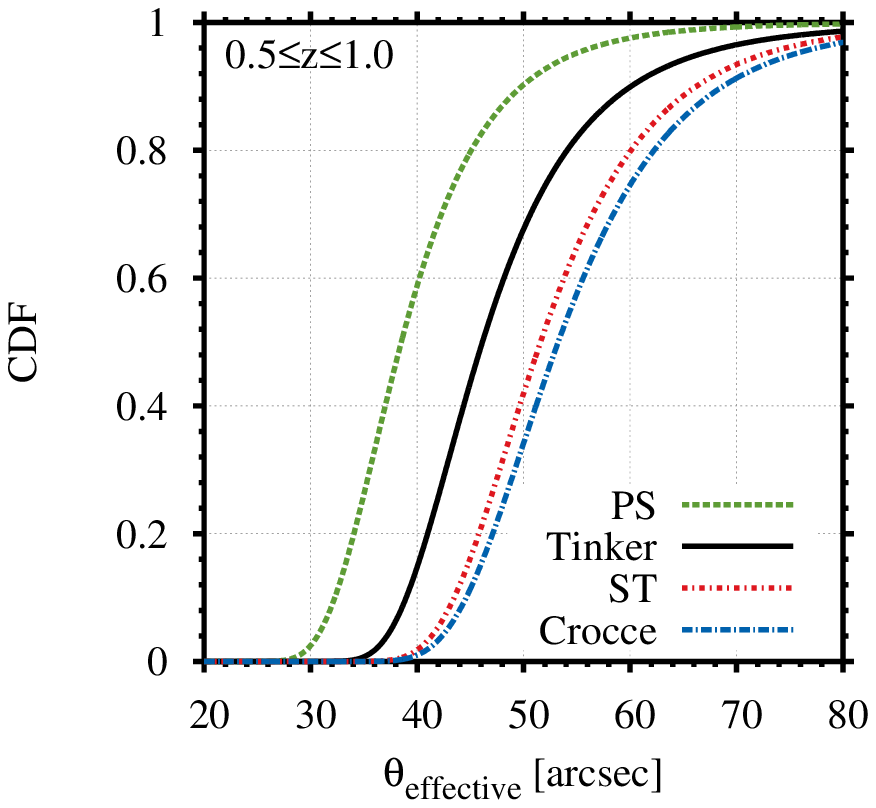}
\includegraphics[width=0.425\linewidth]{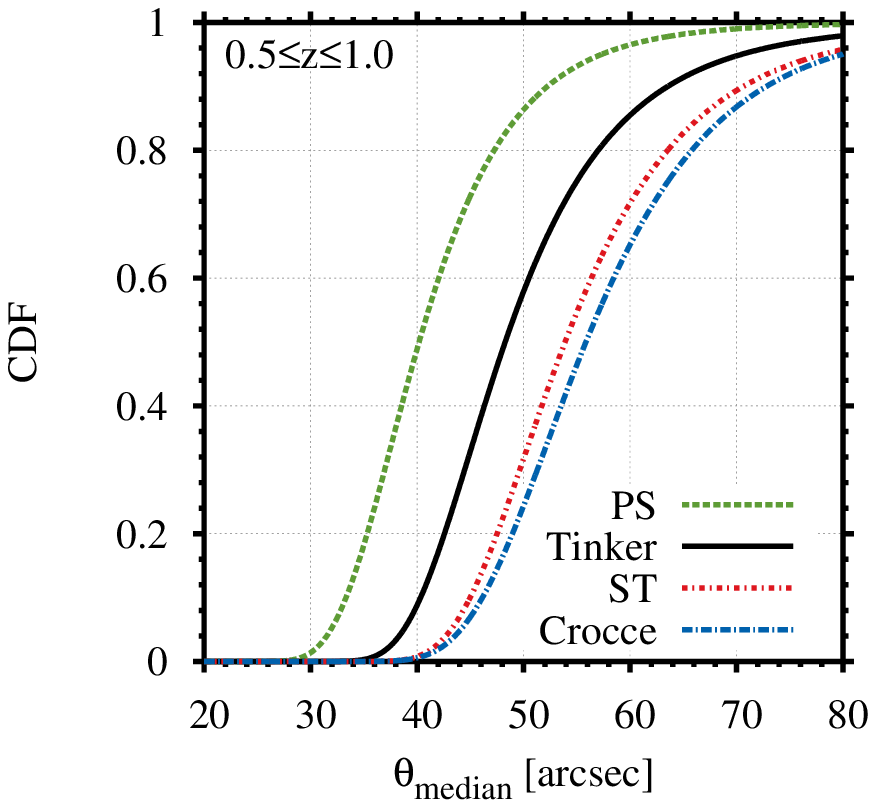}\\
\includegraphics[width=0.425\linewidth]{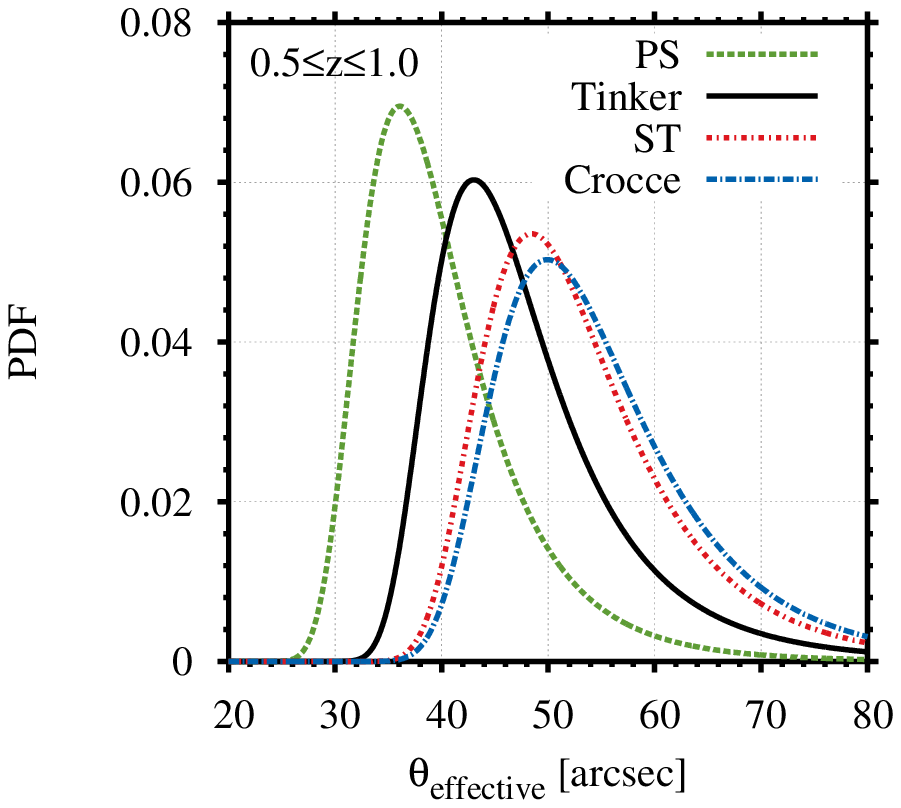}
\includegraphics[width=0.425\linewidth]{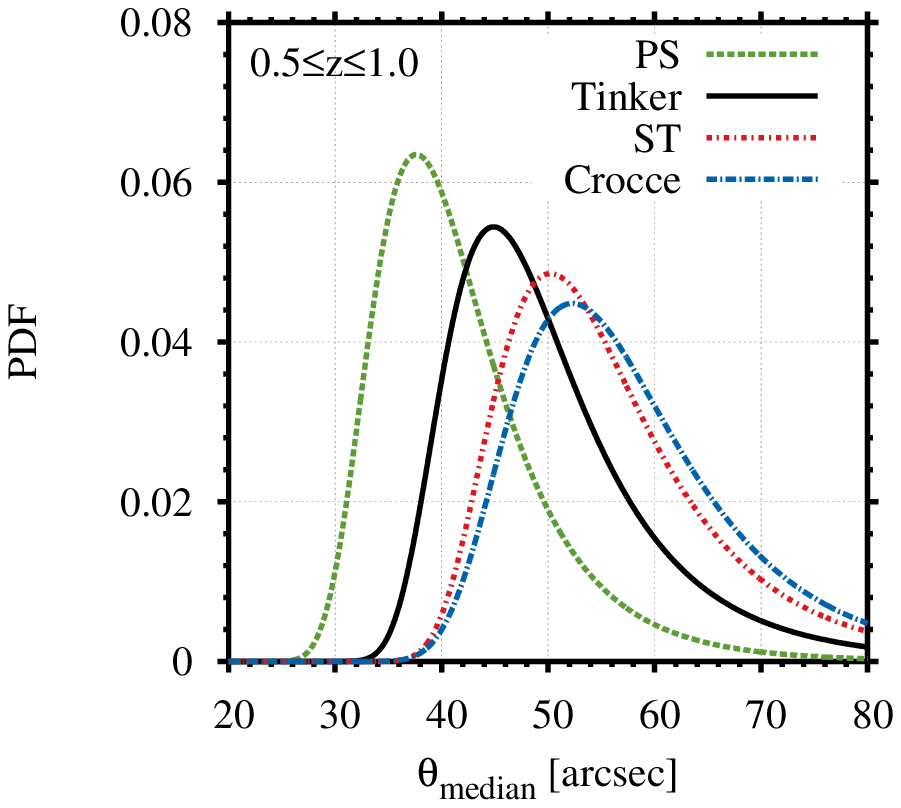}
\caption{CDFs (upper panels) and PDFs (lower panels) of the largest effective (left panels) and 
the median (right panels) Einstein radius for the mass functions of Press-Schechter, 
Tinker, Sheth \& Tormen and Crocce as labelled in the panels. All distributions are based 
on the simulation of $1\,000$ maxima with $M_{\rm lim}=2\times10^{14}\,M_\odot /h$ 
in the redshift interval of $0.5\le z\le 1.0 $ on the full sky.}\label{fig:ER_massfunction}
\end{figure*}

\begin{figure}
\centering
\includegraphics[width=0.95\linewidth]{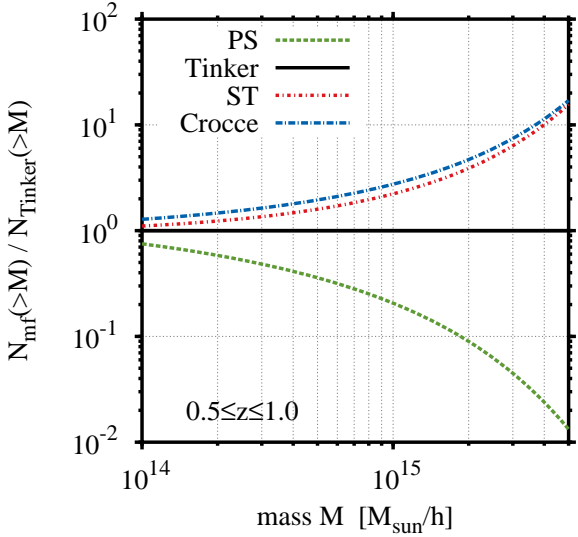}
\caption{Ratio of the number of haloes $N(>M)$ for the different mass functions as 
labelled in the panel, with respect to the Tinker one, assuming a redshift interval of 
$0.5\le z\le 1.0 $ and the full sky.}\label{fig:num_diff_massfnct}
\end{figure}

In order to quantify the influence of different mass functions, we sampled the CDFs of 
the largest Einstein radii for four different mass functions, the \cite{Press1974} 
(PS), \cite{Tinker2008}, \cite{Sheth&Tormen1999} (ST) and the \cite{Crocce2010} mass 
functions. We decided to use the Tinker mass function as a reference because the halo masses are 
defined as spherical overdensities with respect to the mean background density, a definition 
that is closer to theory and actual observations than the friend-of-friend masses that were 
used for the Crocce mass function. We added the \cite{Crocce2010} mass function to our 
analysis because it differs significantly at the high-mass end from other simulations 
\citep{Bhattacharya2011}. The mass function is based on simulations with a box size much 
larger than the horizon scale, which gives more statistics at the high-mass end at the price 
of leaving the realm of the Newtonian approximation. However, \cite{Green2011} argue that 
the Newtonian approximation for N-body simulations might also be valid on super-horizon 
scales. \\
The resulting extreme value distributions for the four different mass functions are presented 
in \autoref{fig:ER_massfunction}, on the basis of the simulation of $1\,000$ maxima with 
$M_{\rm lim}=2\times 10^{14}\,M_\odot /h$ in the redshift interval of $0.5\le z\le 1.0 $ 
on the full sky. The results show that the effect of different mass functions is substantial. 
The CDFs based on the ST and the Crocce mass functions exhibit the strongest difference 
with respect to the PS one. The Tinker results lie in between the two. The modes of the 
distributions, listed in \autoref{tab:modes}, can differ by more than $10\,{\rm arcsec}$ 
for different choices of the mass function, which implies that the inferred occurrence 
probabilities for a given Einstein radius can be substantially different. From the shape 
parameters $\gamma$, it can be inferred that the different mass functions do not strongly 
affect the shape of the distributions. This is also confirmed by the fact that the CDFs shown 
in the upper row of \autoref{fig:ER_massfunction} seem just to be shifted in the $\theta$
-direction.\\ 
The distributions clearly reflect the different behaviour of the mass functions with mass, as 
shown in \autoref{fig:num_diff_massfnct} in the form of the ratio of the number of haloes 
more massive than $M$ with respect to the Tinker case. It can be seen that the ST and Crocce 
mass functions lead to a substantial increase in the number of haloes, particularly at the 
high-mass end, whereas the PS mass function results in far fewer haloes in the redshift 
interval of interest.\\
Considering the strong impact of the mass function on the extreme value distribution, 
it is certainly necessary to improve the accuracy of the mass function at the high-mass end. 
\begin{table}
\centering
\caption{Shape parameter $\gamma$ and the mode of the CDFs of the largest 
effective and median Einstein radius for four different mass functions.}
\begin{tabular}{lcccc} \hline\hline
Mass Function & $\gamma_{\rm eff}$ & Mode($\theta_{\rm eff}$) 
& $\gamma_{\rm median}$ & Mode($\theta_{\rm median}$)  \\
\hline
PS			& $0.091$ & $36.1^{\prime\prime}$ & $0.073$ & $37.6^{\prime\prime}$\\
Tinker	& $0.138$ & $43.0^{\prime\prime}$ & $0.130$ & $44.9^{\prime\prime}$\\
ST 		& $0.089$ & $48.4^{\prime\prime}$ & $0.115$ & $50.2^{\prime\prime}$\\
Crocce	& $0.084$ & $49.9^{\prime\prime}$ & $0.067$ & $52.3^{\prime\prime}$\\
\hline
\end{tabular}\label{tab:modes}
\end{table}
\subsection{The impact of triaxiality}\label{sub:triaxiality}
The triaxiality of the lensing haloes has a substantial impact on the distribution of 
the maxima. For instance, a very elongated halo that is directed along the line of sight 
can lead to a highly concentrated, projected surface mass-density profile, which causes 
a large tangential critical curve \citep[see e.g.][]{Oguri2003, Dalal2004, 
Meneghetti2007, Meneghetti2010}. When sampling axis ratios from \autoref{eq:p_a} 
and \autoref{eq:p_ab}, particularly small values of the sampled axis ratios will potentially 
propagate into extreme strong-lensing events. Since the empirical fits from JS02 are only 
based on few data points in this regime, it is unclear how reliable the fitted axis-ratio 
distributions are.

\begin{figure*}
\centering
\includegraphics[width=0.425\linewidth]{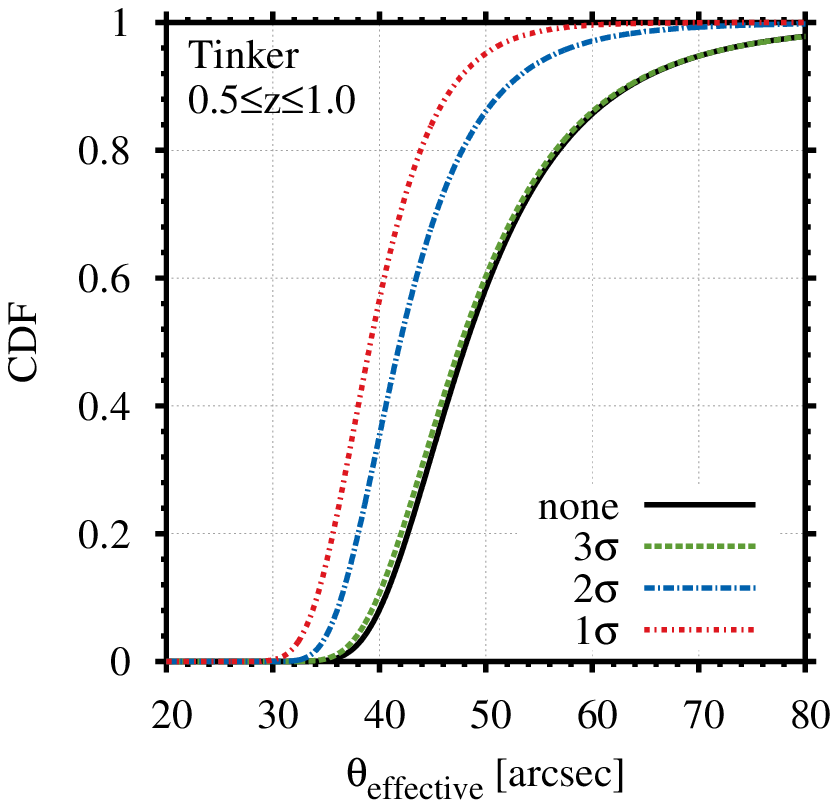}
\includegraphics[width=0.425\linewidth]{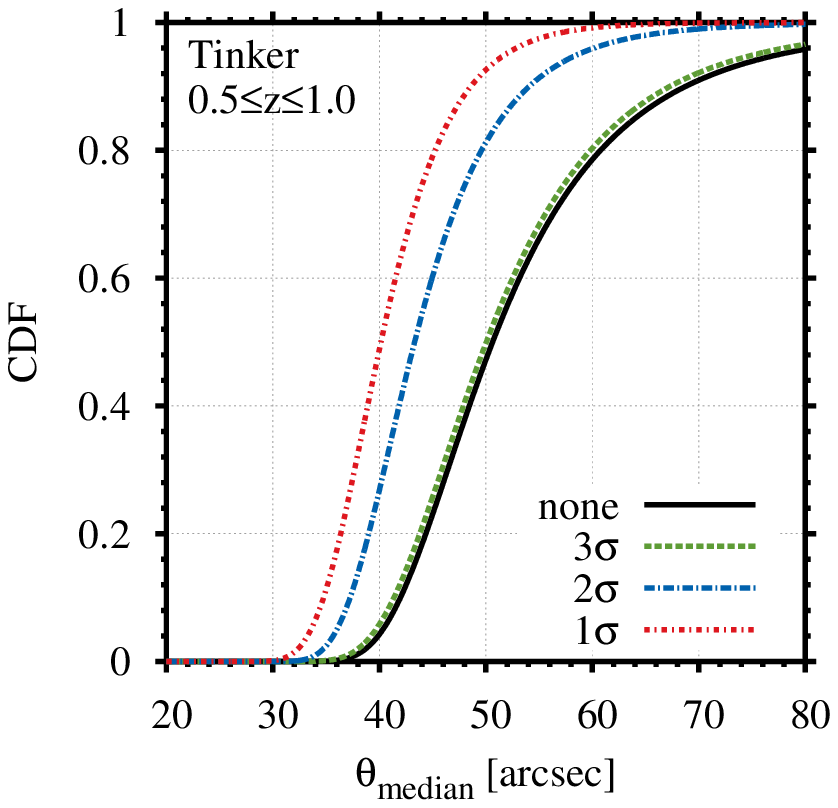}\\
\includegraphics[width=0.425\linewidth]{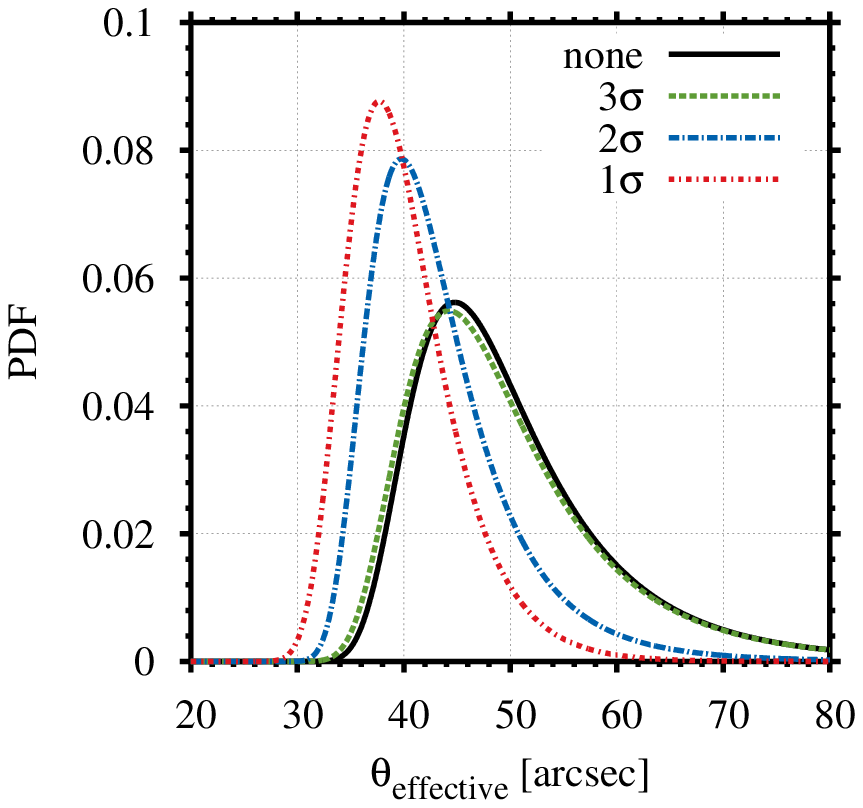}
\includegraphics[width=0.425\linewidth]{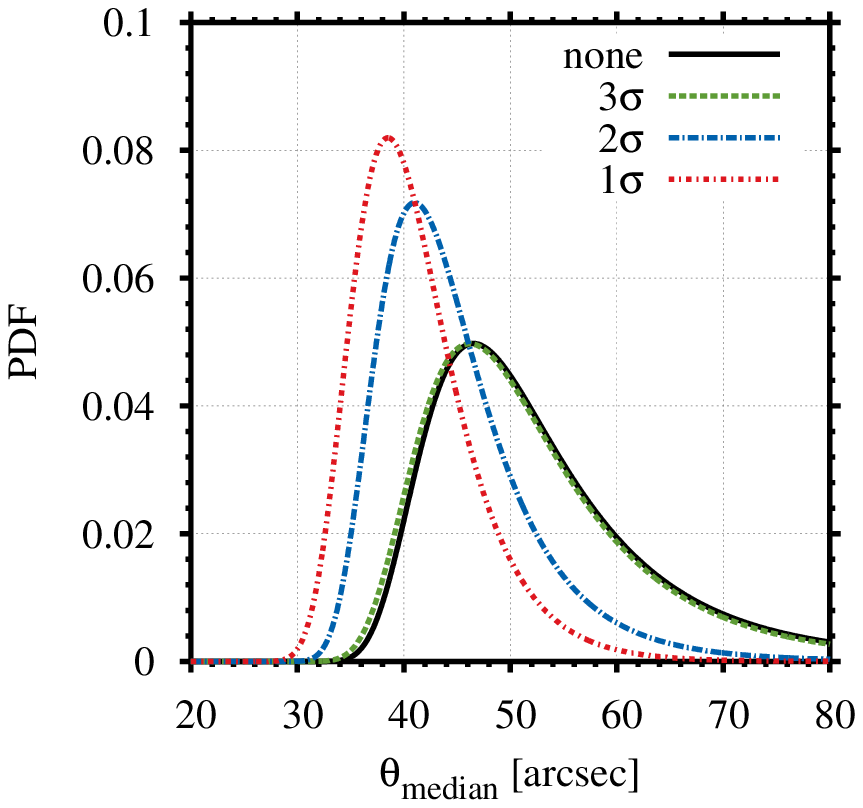}
\caption{CDFs (upper panels) and PDFs (lower panels) of the largest effective (left panels) and median 
(right panels) Einstein radius for cut-offs of the axis-ratio distribution as labelled 
in the figure. All distributions are based on the \cite{Tinker2008} mass function and the simulation of $1\,000$ 
maxima with $M_{\rm lim}=2\times10^{14}\,M_\odot /h$ in the redshift interval of $0.5\le z\le 1.0 $ 
on the full sky.}\label{fig:ER_cutoff}
\end{figure*}

In order to study the impact of this uncertainty, we introduced a cut-off in the distribution 
from \autoref{eq:p_a} to remove extreme axis ratios. We cut off the distribution of 
the scaled axis ratios at different confidence levels $n$, according to
\begin{equation}
a^{\rm cut}_{\rm sc}=0.54 - n\sigma_{\rm s},
\end{equation}
selecting values of $1\sigma$, $2\sigma$, $3\sigma$ and comparing them to the case without 
any cut-off. To do so, we MC--simulate the distributions of the largest Einstein radius for the 
different cut-offs based on $1\,000$ maxima, assuming $M_{\rm lim}=2\times 10^{14}
\,M_\odot /h$ and the Tinker mass function. The results are presented in \autoref{fig:ER_cutoff}, 
which shows that the impact of the tail of the axis ratio distribution on the 
extreme value distribution is substantial. In comparison to the impact of the different mass 
functions that lead to a shift of the CDF, the cut-off value strongly affects the mode as well 
as the shape of the CDFs, as can be inferred from the shape parameters listed in 
\autoref{tab:cutoff}. For the $1\sigma$ cut-off, the shape parameter becomes negative. 
Consequently, the CDF of the largest Einstein radius is then described by a Weibull (Type III) 
distribution, indicating that the underlying distribution is bounded from above. For decreasing 
values of the cut-off (less extreme axis ratios), the CDF steepens, which corresponds to the 
fact that a given observed Einstein radius appears to be less likely to exist.

\begin{figure*}
\centering
\includegraphics[height=5.1cm]{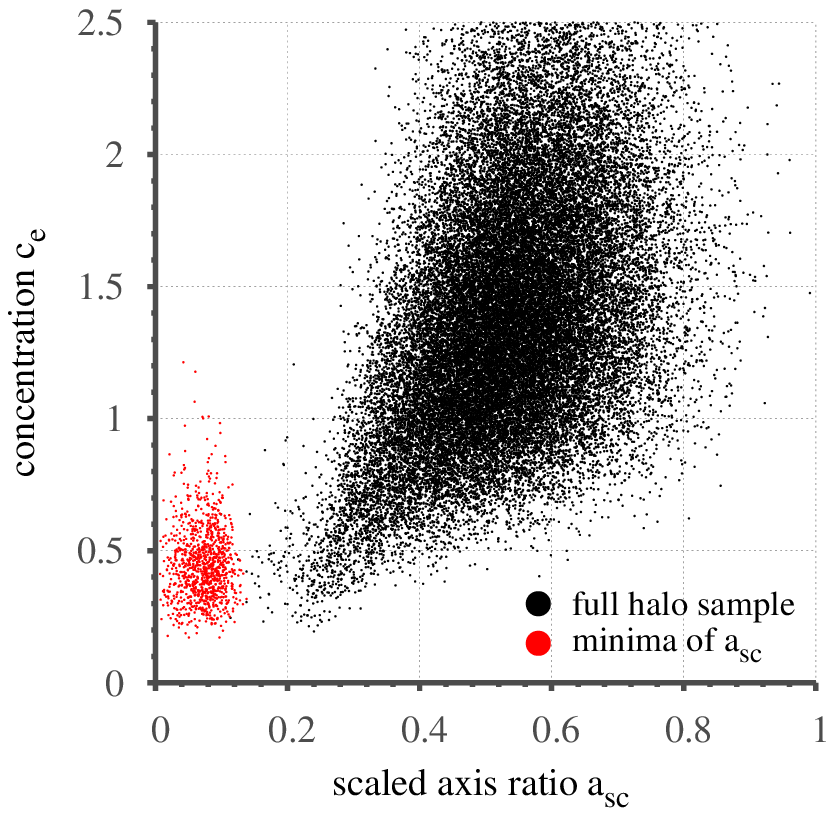}
\includegraphics[height=5.1cm]{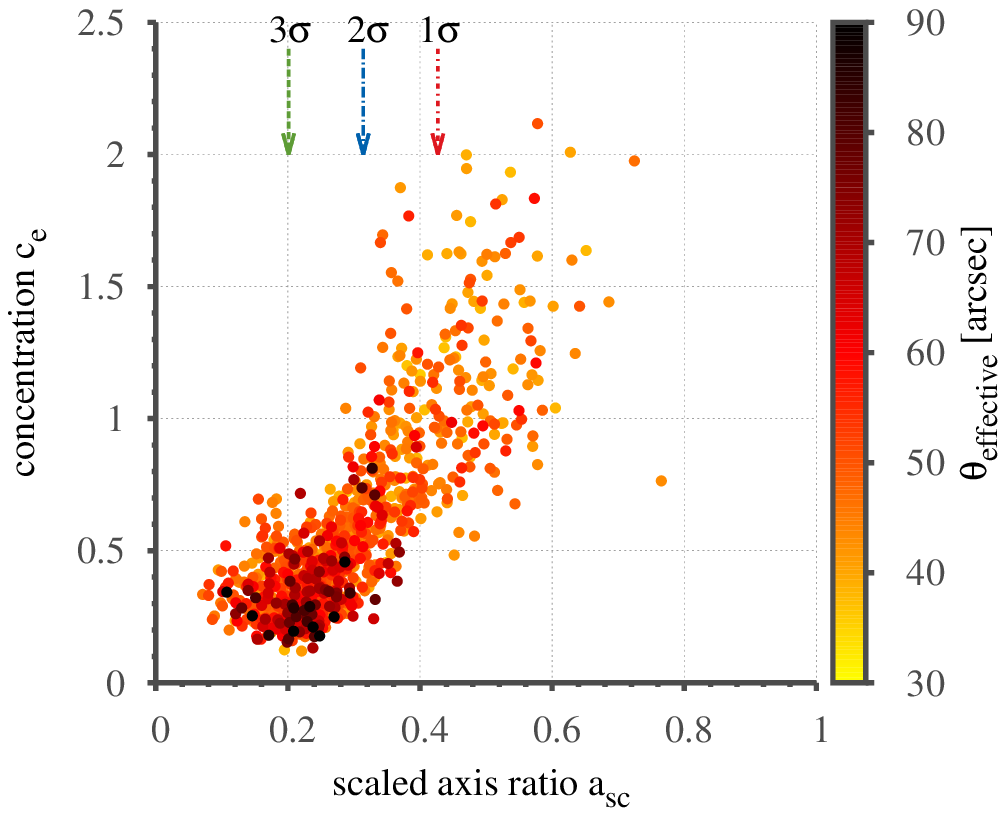}
\includegraphics[height=5.1cm]{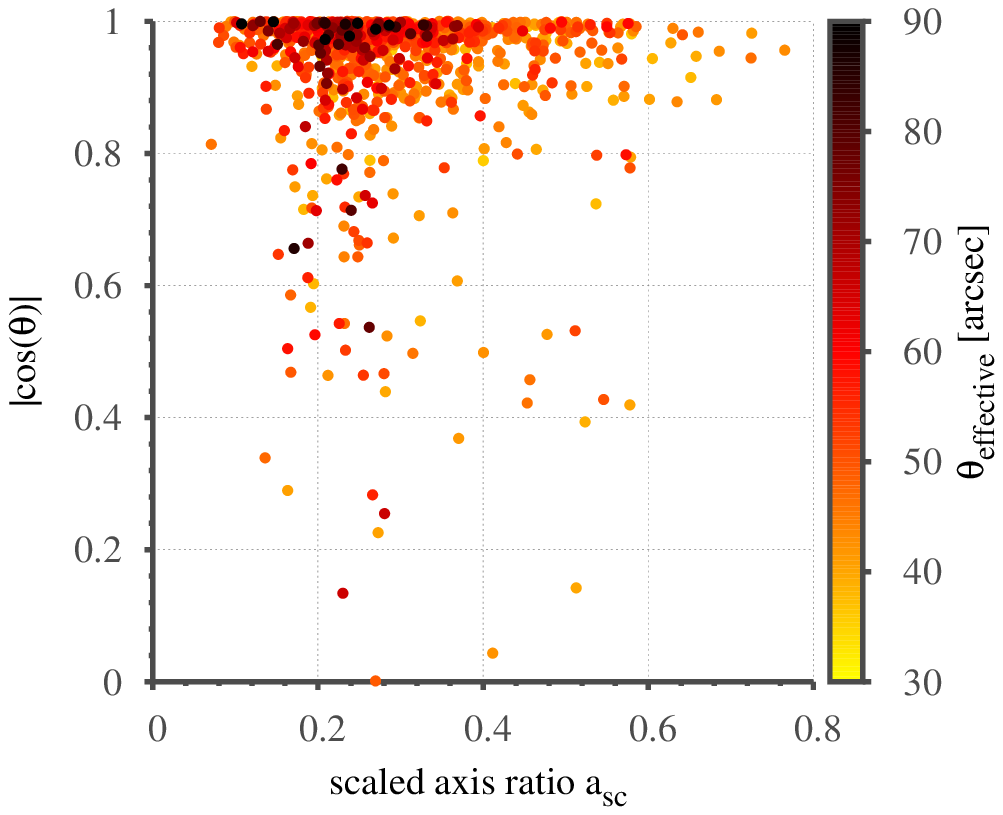}
\caption{The left-hand and centre panels show the distribution in scaled axis ratio $a_{\rm sc}$ 
and concentration $c_{\rm e}$ of sampled haloes, according to different 
selection criteria. The left-hand panel shows the distribution of a full halo sample of a 
single realisation (black dots) and the sample of the minima in $a_{\rm sc}$ from $1\,000$ 
realisations (red dots). The centre panel shows the distribution of the haloes that give rise 
to the largest effective Einstein ring (encoded in the colourbar) based on $1\,000$ realisations. 
The small arrows denote the indicated cut-offs identical to \autoref{fig:ER_cutoff}. The 
right-hand panel shows the distribution of the same maxima with respect to the scaled axis 
ratio and the alignment $|\cos{\theta}\,|$, where $\theta$ is the angle between the major axis 
of each halo with respect to the line of sight.}\label{fig:asc_ce_distribution}
\end{figure*}

In order to understand better what impact the triaxiality has on the sample of the largest Einstein 
radii, we study the distribution of the sampled haloes in scaled axis ratio $a_{\rm sc}$ and 
concentration $c_{\rm e}$ based on different selection criteria. We are interested in the question 
whether or not the largest Einstein radii always stem from very extreme axis ratios. In the 
left-hand panel of \autoref{fig:asc_ce_distribution}, we show the distribution of the full halo sample 
of a single realisation (black dots) together with the distribution of the minima of $a_{\rm sc}$ 
based on $1\,000$ all-sky realisations. As expected, the highly elongated haloes exhibit small values 
of the concentration parameter and typical values for the minima scatter around 
$\bar{a}_{\rm sc}^{\rm min}\approx 0.08$. Now we compare this distribution to the one based 
on the largest effective Einstein radii shown in the centre panel of 
\autoref{fig:asc_ce_distribution}. It is evident that the largest Einstein radii stem by no means 
exclusively from the haloes with a minimum of $a_{\rm sc}$, but from a rather broad range 
of $a_{\rm sc}$. Thus, the largest Einstein radii either stem from lowly concentrated very elongated 
haloes or from less elongated but higher concentrated ones. However, the largest maxima (the 
dark red to black dots in the centre panel of \autoref{fig:asc_ce_distribution}) are confined to 
the region of small $a_{\rm sc}$ and small $c_{\rm e}$. This fact explains the strong 
impact of the cut-off in the scaled axis-ratio distribution on the shape of the CDF of maxima 
because the cut-off effectively removes the largest maxima.

Due to the limited knowledge of the statistics of extremely small axis ratios, it is not possible to 
clearly define a proper choice of the cut-off (if present) until the triaxiality distributions of large 
halo samples (covering the largest cluster masses) are studied in numerical simulations. In the 
study of JS02 (see their Figure 9), scaled axis ratios below $\sim 0.2$ were not found for any 
of the studied redshifts. The value of $a^{\rm min}_{\rm sc}\sim 0.2$ corresponds to the cut-off 
on the $3\sigma$ level. It should be noted that in general $a_{\rm sc}$ also depends on the 
underlying cosmology, as can be seen from \autoref{eq:a_sc}. Adopting such a cut-off value would 
mean that the resulting CDFs of the maxima are very close to the one without any cut-off, as can be 
seen from \autoref{fig:ER_cutoff}. Of course, \autoref{fig:asc_ce_distribution} indicates that the 
impact of triaxiality should always be discussed together with that of the concentration. 
\begin{table}
\centering
\caption{Shape parameter $\gamma$ and the mode of the CDFs for the largest effective and 
median Einstein radius for different cut-offs in the axis-ratio distribution.}
\begin{tabular}{lccccc} \hline\hline
Cut-off & $a^{\rm cut}_{\rm sc}$ & $\gamma_{\rm eff}$ & Mode($\theta_{\rm eff}$) 
& $\gamma_{\rm median}$ & Mode($\theta_{\rm median}$)  \\
\hline
$1\sigma$	& $0.427$ & $-0.014$ & $37.7^{\prime\prime}$ & $-0.001$ & $38.5^{\prime\prime}$\\
$2\sigma$	& $0.314$ & $0.098$ & $39.8^{\prime\prime}$ & $0.079$ & $41.0^{\prime\prime}$\\
$3\sigma$ 	& $0.201$ & $0.149$ & $44.2^{\prime\prime}$ & $0.151$ & $46.2^{\prime\prime}$\\
none				& -- & $0.151$ & $44.8^{\prime\prime}$ & $0.185$ & $46.5^{\prime\prime}$\\
\hline
\end{tabular}\label{tab:cutoff}
\end{table}
\begin{figure*}
\centering
\includegraphics[width=0.425\linewidth]{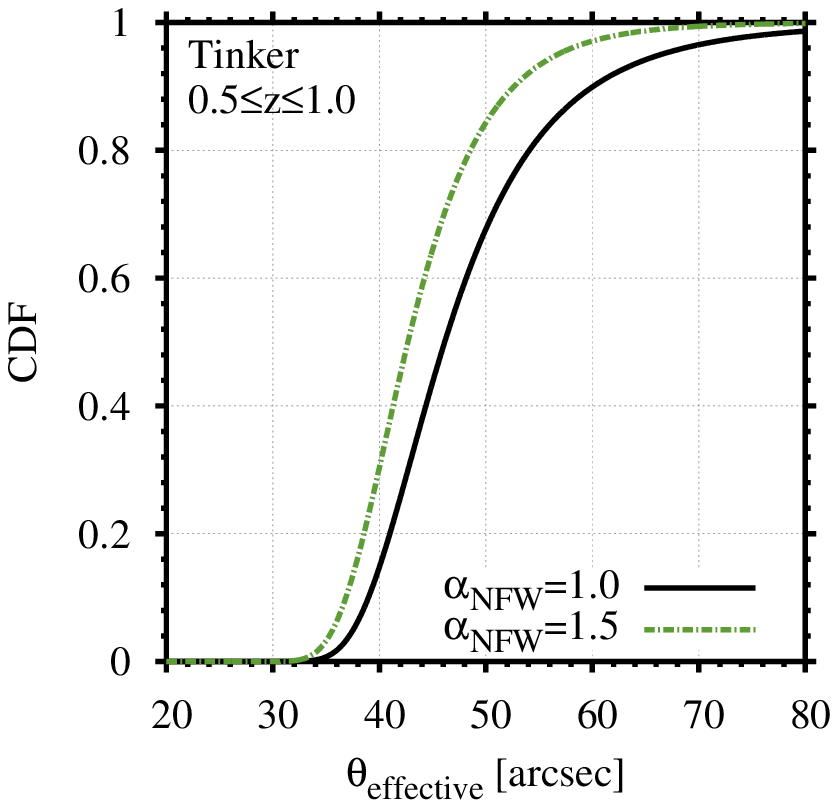}
\includegraphics[width=0.425\linewidth]{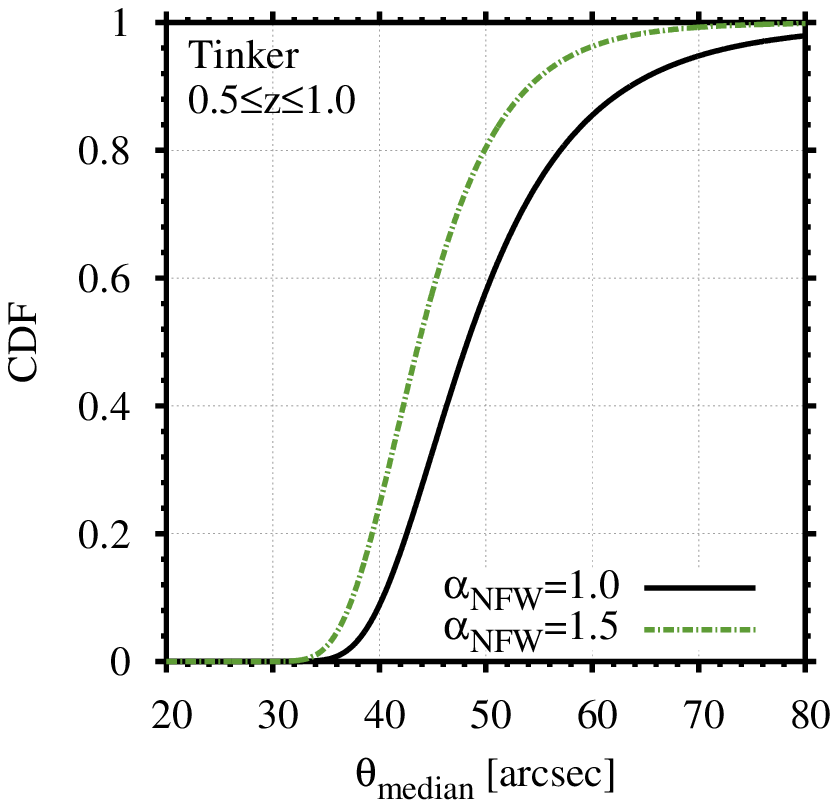}\\
\includegraphics[width=0.425\linewidth]{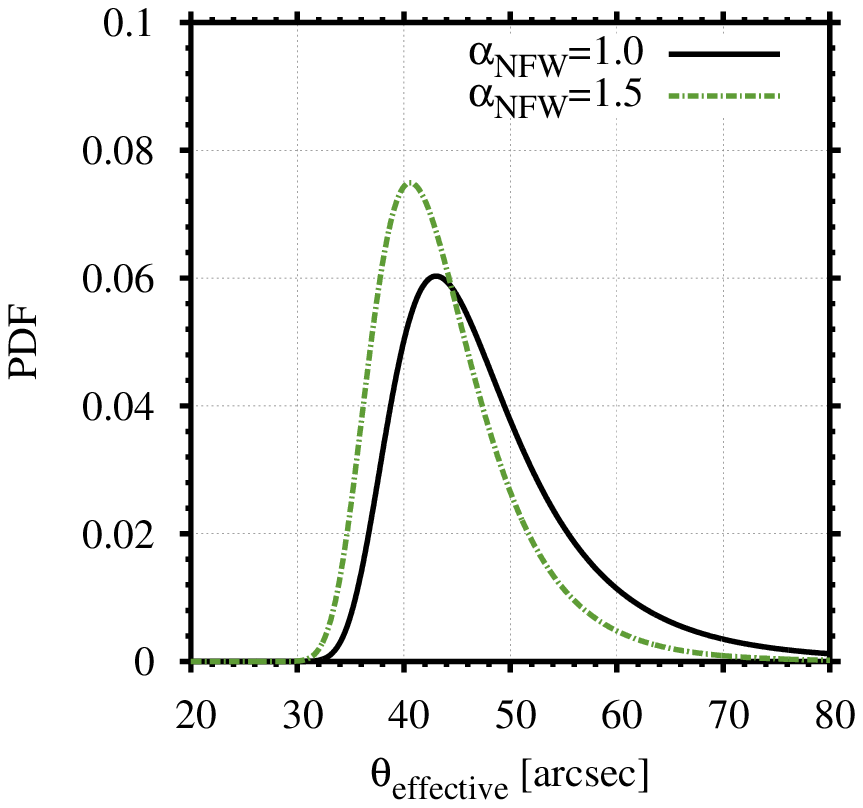}
\includegraphics[width=0.425\linewidth]{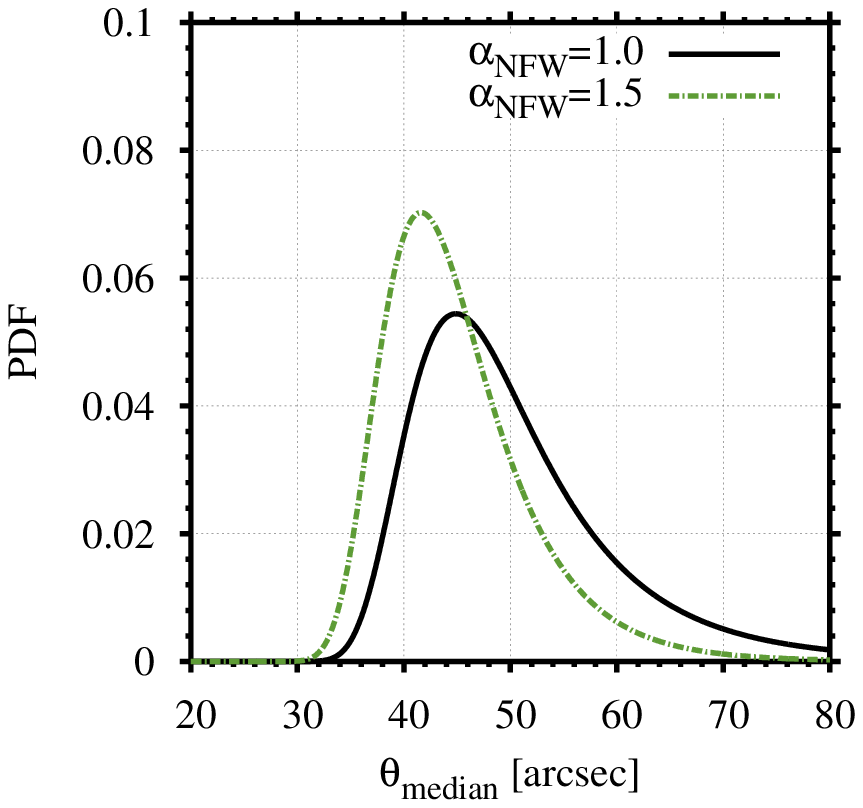}
\caption{CDFs (upper panels) and PDFs (lower panels) of the largest effective (left panels) and 
median (right panels) Einstein radius for different values of the inner slope of the density profile 
$\alpha_{\rm NFW}=(1.0, 1.5)$ as labelled in the figure. All distributions are based on the 
\cite{Tinker2008} mass function and the simulation of $1\,000$ maxima with $M_{\rm lim}=
2\times10^{14} \,M_\odot /h$ in the redshift interval of $0.5\le z\le 1.0 $ on the full sky.}
\label{fig:ER_innerslope}
\end{figure*}
\subsection{The impact of the inner slope and the $c$--$M$ relation}
In order to study the effect of the inner slope $\alpha_{\rm NFW}$ of the density 
profile, we sampled the distribution of the maxima for two different values, 
$\alpha_{\rm NFW}=(1.0, 1.5)$, using the approach discussed at the end of 
\autoref{sec:singleHaloes}. The resulting distributions are presented in 
\autoref{fig:ER_innerslope} for $\alpha_{\rm NFW} = 1.0$ (black solid line) and 
$\alpha_{\rm NFW} = 1.5$ (green dashed-dotted line). The distribution for the 
steeper inner density profile is shifted to smaller Einstein radii, confirming the 
findings reported in \cite{Oguri&Keeton2004} on p.~669. On average, steeper 
density profiles lead to slightly larger Einstein radii \citep[see e.g.][]{OguriPhD2004}. 
However, the distribution of the maxima is particularly sensitive to the triaxiality 
and the orientation of the halo along the line of sight. For the corresponding small 
axis ratios, the shallower density profile contributes stronger due to the projection 
effect\footnote{Note that we do not truncate the density profile.} if a very elongated, 
extended halo is aligned along the line of sight.

\begin{figure*}
\centering
\includegraphics[width=0.425\linewidth]{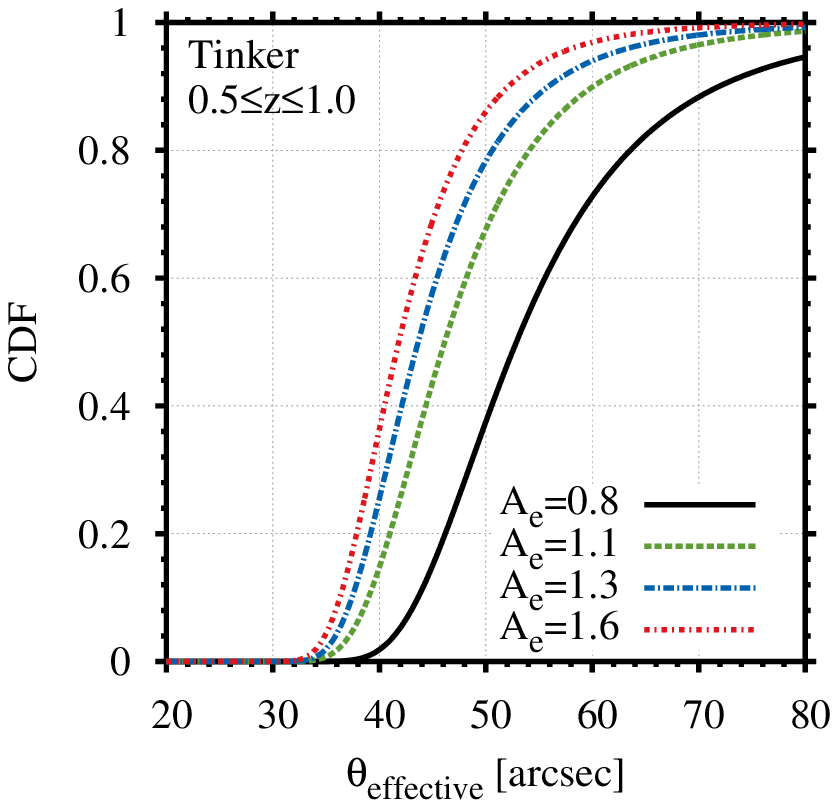}
\includegraphics[width=0.425\linewidth]{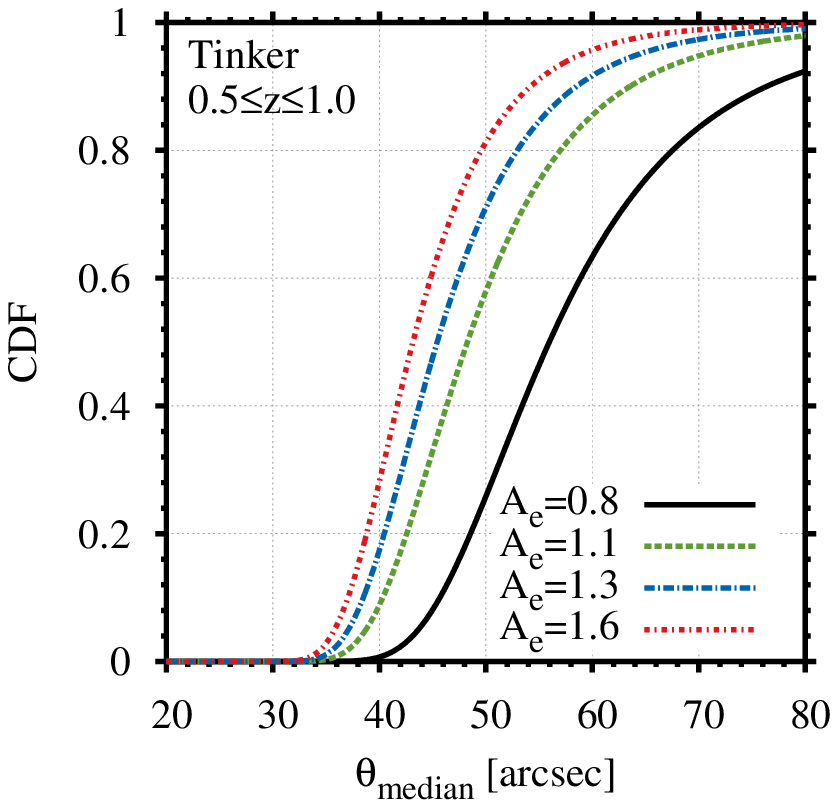}\\
\includegraphics[width=0.425\linewidth]{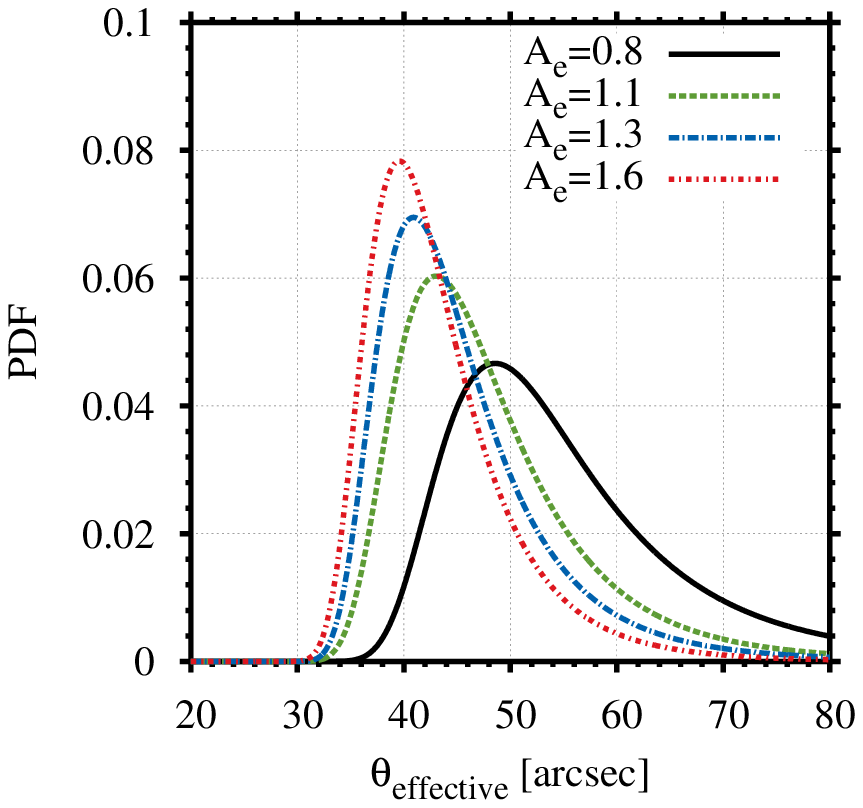}
\includegraphics[width=0.425\linewidth]{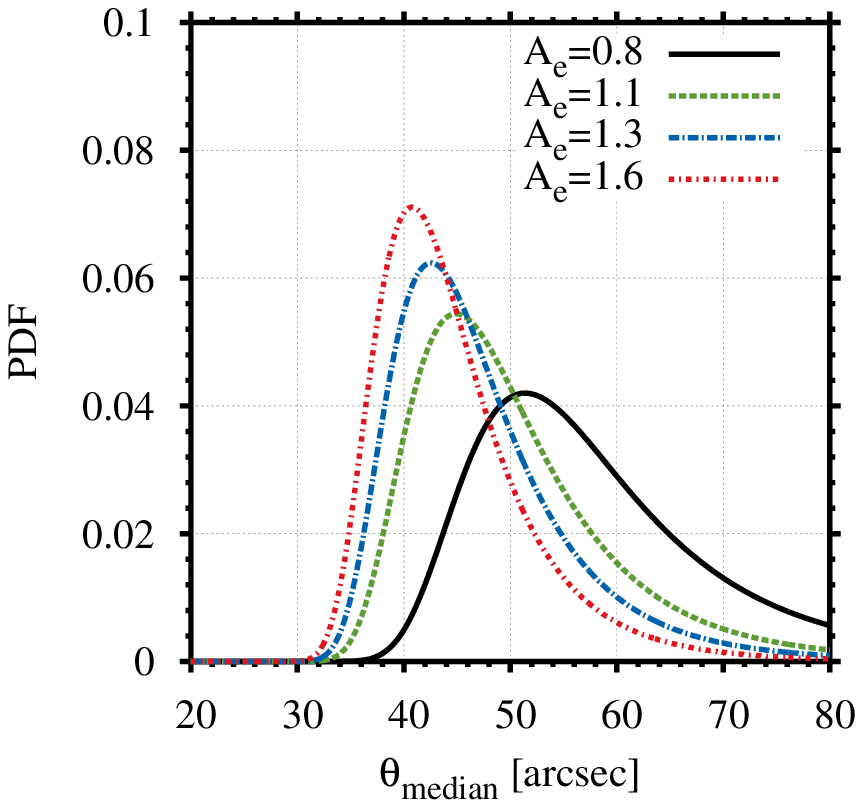}
\caption{CDFs (upper panels) and PDFs (lower panels) of the largest effective (left panels) and median 
(right panels) Einstein radius for different values of the normalisation parameter $A_{\rm e}$
of the mass-concentration relation as labelled in the figure. All distributions are based on the 
\cite{Tinker2008} mass function and the simulation of $1\,000$ maxima with $M_{\rm lim}=2\times10^{14}
\,M_\odot /h$ in the redshift interval of $0.5\le z\le 1.0 $ on 
the full sky.}\label{fig:ER_cm_relation}
\end{figure*}

In addition to the previously discussed effects, the $c$--$M$ relation naturally 
impacts significantly on the distribution of the largest Einstein radii. In order to 
study the impact, we mimic a variation in the $c$--$M$ relation by computing 
the distributions for different values of the normalisation parameter, $A_{\rm e}$, 
in \autoref{eq:ce} for the mean concentration. Here, assuming a fixed $a_{\rm sc}$, 
smaller $A_{\rm e}$ correspond to smaller values of the mean concentration, and 
thus it is more likely that haloes have a smaller $c_{\rm e}$. 
Following \cite{Oguri2003}, we vary the value of $A_{\rm e}$ between $0.8$ and 
$1.6$, with $A_{\rm e}=1.1$ being the $\Lambda$CDM standard value, and present 
the resulting distributions in \autoref{fig:ER_cm_relation}. It can be seen that the 
distributions with a larger value of $A_{\rm e}$ are shifted to smaller Einstein radii 
with respect to the standard choice of $A_{\rm e}=1.1$. Vice versa, the 
distribution for $A_{\rm e}=0.8$ is shifted to larger values, confirming the findings 
from the previous section and \autoref{fig:asc_ce_distribution} that the largest maxima 
stem from haloes with small $a_{\rm sc}$ and $c_{\rm e}$. Thus, lowering the mean 
concentration with a smaller value of $A_{\rm e}$ naturally results in shifting the CDF 
of the maxima to larger values. In this sense, the results presented in 
\autoref{fig:ER_cm_relation} reflect the high sensitivity of the maxima to particularly 
small axis ratios, which are connected to small concentrations (larger scale radii) and, 
hence, more extended haloes. When aligned along the line of sight, such a system 
will give rise to a particularly large Einstein radius. In fact, the right-hand panel of 
\autoref{fig:asc_ce_distribution} shows that the vast majority of maxima of the Einstein 
radius are very well aligned along the line of sight. 

With the results of this section, we mainly want to emphasize that there are many 
uncertainties having a strong impact on the statistics of Einstein radii. 
In particular, we would like to underline the need to improve the statistics of triaxiality 
for extreme axis ratios and the closely related uncertainties stemming from projection effects.
\subsection{Other important effects}
The most important effect that will strongly influence the distribution of the maxima is the inclusion of 
dynamical mergers as discussed in \cite{Redlich2012}. In particular, mergers perpendicular to the 
line of sight can cause a strong elongation of the critical curve. Closely related and to some extent equivalent, 
the inclusion of substructures is expected to have a similar effect. Also, the brightest cluster galaxy (BCG) 
can lead to an increase of the strong-lensing effect \citep{Puchwein2005}, but to a lesser extent than the 
previously discussed effects. We decided to omit mergers in this study, since their inclusion is computationally 
expensive and, as the next section shows, they are not required to explain the presence of the largest known 
Einstein radius.
\section{MACS J0717.5+3745 -- A case study}\label{sec:MACSJ0717}
After the statistical study of the extreme value distributions of the largest Einstein radius and the 
different effects that influence them, we use the inferred distribution to assess the occurrence 
probability of the Einstein radius of MACS~J0717.5+3745, which has the largest currently known 
observed critical curve. In addition, we also discuss in the second part of this section, what mass 
MACS~J0717.5+3745 would need to have in order to be considered to be in significant tension 
with the standard $\Lambda$CDM model.
\subsection{The cluster MACS J0717.5+3745}
The X-ray luminous galaxy cluster MACS J0717.5+3745, independently observed at redshift 
$z_{\rm obs}=0.546$ by the MACS survey \citep{Ebeling2001, Ebeling2007} and as a host 
of a diffuse radio source by \cite{Edge2003}, is a quite remarkable system. The cluster is 
connected to a $4\,{\rm Mpc}$ long large-scale filament \citep{Ebeling2004} that leads into 
the cluster and exhibits ongoing merging activity \citep{Ma2008}. Furthermore, this cluster 
possesses the most powerful known radio halo \citep{Bonafede2009} and has also been 
observed \citep{LaRoque2003, Mroczkowski2012} via the Sunyaev Zeldovich effect \citep{Sunyaev1972, 
Sunyaev1980}. A strong-lensing analysis \citep{Zitrin2009, Zitrin2011a} of this highly 
interesting system revealed that, with $\theta_{\rm eff}=55\pm3\,{\rm arcsec}$ for an estimated 
source redshift of $z_{\rm s}\sim 2.5$, the effective Einstein radius is the largest known 
at redshifts $z>0.5$. Also, the mass enclosed by this critical curve is with $(7.4\pm 0.5)\times 
10^{14}\,M_\odot$ very large, indicating that this cluster might qualify to be among the most 
massive known galaxy clusters at $z>0.5$. The recent strong-lensing analysis by \cite{Limousin2011} 
reports a higher redshift of $z_{\rm s}\sim 2.96$ for the primary lensed system, which would lower 
the size of the Einstein radius as well as the overall mass estimate.
\subsection{The probability of occurrence of the critical curve}
\begin{figure*}
\centering
\includegraphics[width=0.495\linewidth]{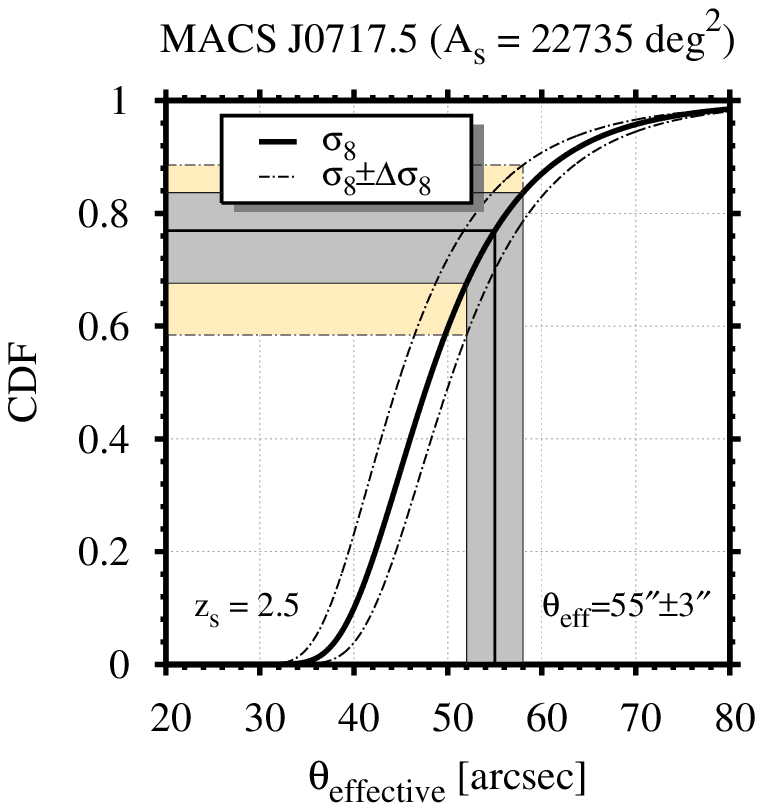}
\includegraphics[width=0.495\linewidth]{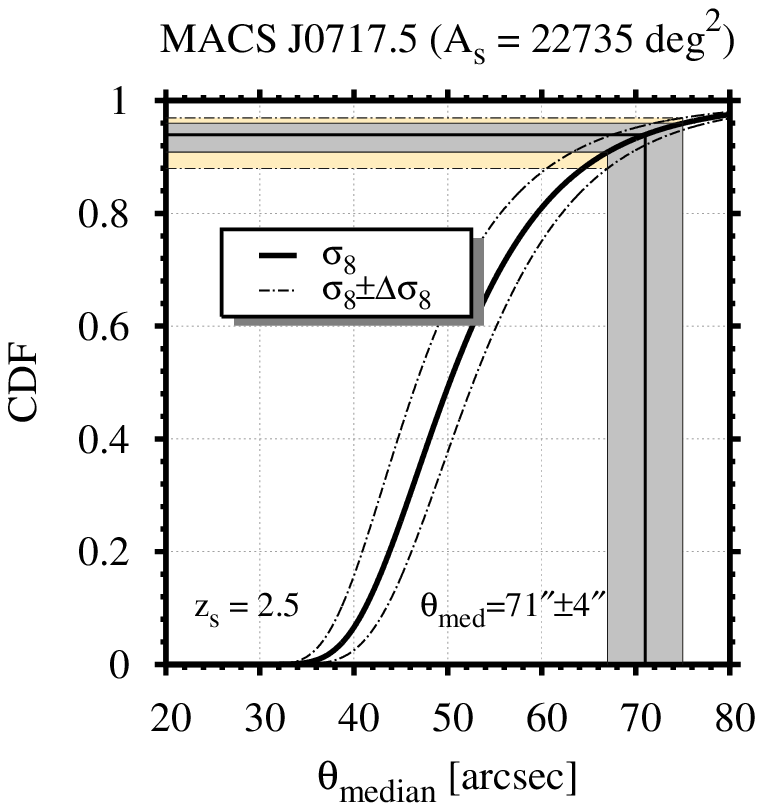}\\
\includegraphics[width=0.495\linewidth]{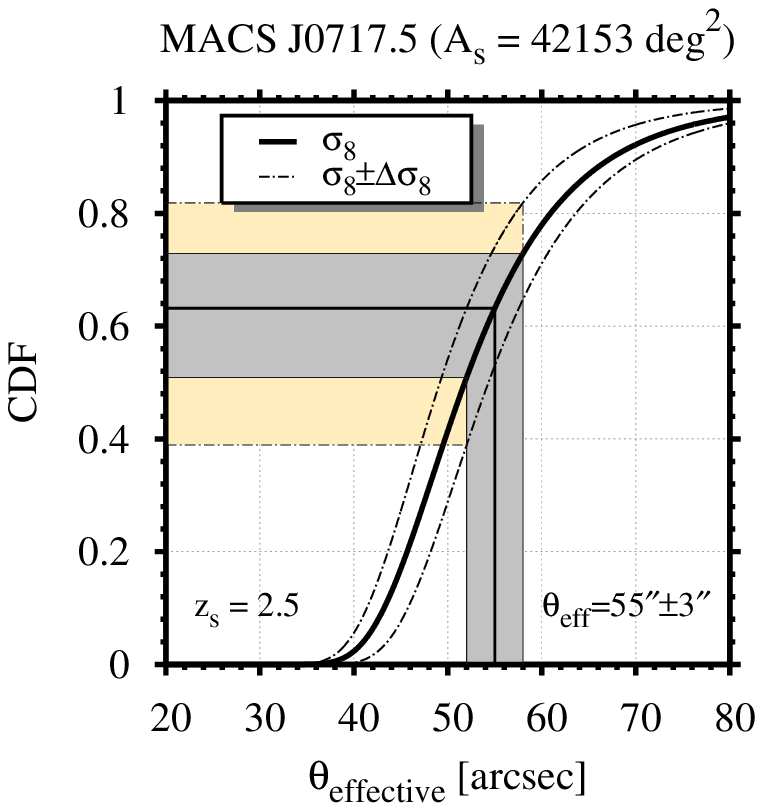}
\includegraphics[width=0.495\linewidth]{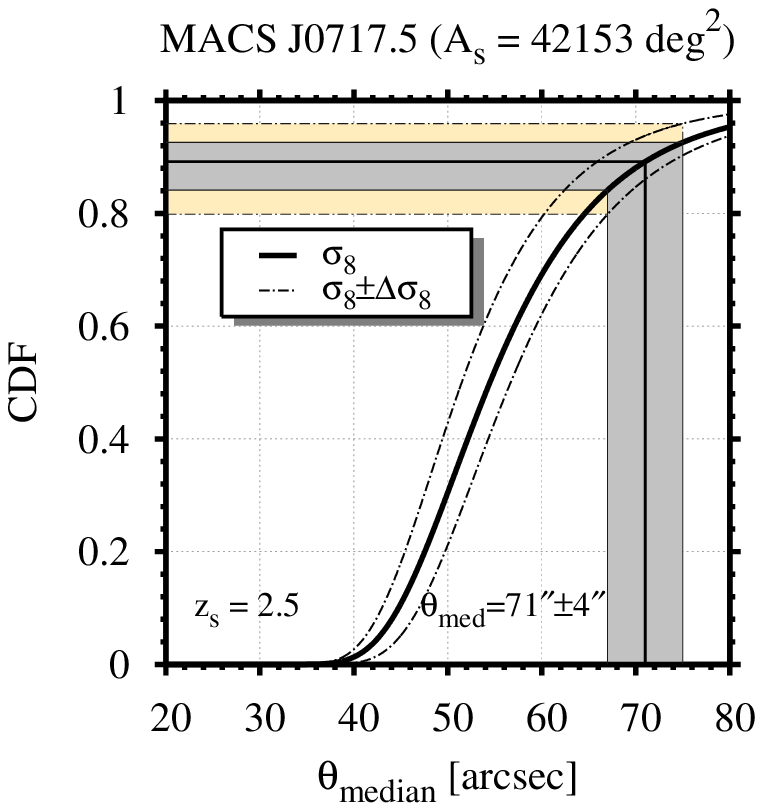}
\caption{CDFs of the largest effective (left panel) and median (right panel) Einstein radius, assuming 
a redshift interval of $0.5\le z\le 1.0 $, a source redshift of $z_{\rm s}=2.5$ and the nominal MACS 
survey area (upper panels) as well as the full sky (lower panels). Both distributions are based on the 
\cite{Tinker2008} mass function and the simulation of $1\,000$ maxima with $M_{\rm lim}=2\times
10^{14}\,M_\odot /h$ on the full sky. The dashed-dotted lines, together with the yellow shaded area, 
illustrate the impact of the uncertainty in the\textit{ WMAP7} value of $\sigma_8$ on the CDFs and the 
grey shaded area denotes the uncertainty in the measurement of the Einstein radius.}
\label{fig:MACS_comparison}
\end{figure*}

In order to assess the occurrence probability of the observed effective Einstein radius of 
MACS~J0717.5+3745, we compute the CDF of the largest Einstein radius for a redshift interval 
of $0.5\le z\le 1.0$, based on the Tinker mass function, a source redshift of $z_{\rm s}=
2.5$ for the nominal MACS survey area $(A_{\rm s}=22\,735\,{\rm deg}^2)$, and the 
full sky $(A_{\rm s}=41\,153\,{\rm deg}^2)$. We decided to use the conservative cut-off  $a^{\rm cut}_{\rm sc}=0.249$, 
corresponding to the inclusion of $99$ per cent of the possible axis ratios from the distribution 
in \autoref{eq:p_a}. In doing so, we cut off the distribution above the most likely 
minimum of $a^{\rm min}_{\rm sc}=0.2$. Thus the CDF will be steeper, resulting in a 
more conservative estimate of the occurrence probability of a given Einstein radius. In comparison 
to the previously discussed distributions that assumed a $z_{\rm s}=2.0$, the higher source 
redshift will shift the distribution to larger Einstein radii due to the modified lensing geometry, 
as discussed in \cite{Oguri2009}.\\
Like mass function, the uncertainty in the normalisation of the matter power spectrum, 
$\sigma_8$, will also influence the distribution of the largest Einstein radius, since it influences the 
number of haloes from which the maxima will be drawn. In order to incorporate the uncertainty 
in the measured $\sigma_8$, we also computed the distributions for the upper and lower 
$1\sigma$ limits $\sigma_8=0.811\pm 0.023$ of the \textit{WMAP7}+\textit{BAO}+\textit{SNSALT} 
dataset \citep{Komatsu2011}. The results are shown in \autoref{fig:MACS_comparison} for the 
effective (left-hand panel) and the median (right-hand panel) Einstein radius, where the dashed-dotted 
lines show the CDFs for the upper and lower allowed  value of $\sigma_8$. The upper and the 
lower panels of \autoref{fig:MACS_comparison} show the distributions for the nominal MACS survey area 
and for the full sky, respectively. The grey shaded area 
illustrates the uncertainty due to the accuracy of the measurement of the Einstein radius 
itself and the yellow shaded region depicts the uncertainty due to the allowed range of $\sigma_8$. 
For the MACS survey area and the effective Einstein radius $\theta_{\rm eff}$, we find an 
occurrence probability of $\sim(16-32)$ per cent based on the uncertainty in $\theta_{\rm eff}$ alone. 
When the additional uncertainty of the precision of $\sigma_8$ is included, this range widens to 
$\sim(11-42)$ per cent. The CDF for the median Einstein radius leads to smaller occurrence probabilities 
of $\sim(4-9)$ per cent and $\sim(3-12)$ per cent; however, the median Einstein radius is more sensitive 
to the individual structure of the system. Thus, we decide to base our study on the more conservative 
choice of the effective Einstein radius. When the survey area is extended to the full sky, the CDFs are shifted 
to larger values of the Einstein radius. As a result, the occurrence probabilities for a given observation will 
increase. In the case of MACS J0717.5+3745, we find $\sim(18-61)$ per cent for the effective Einstein 
radius and $\sim(4-20)$ per cent for median Einstein radius when taking the uncertainty of $\sigma_8$ 
into account.

From the ranges of occurrence probabilities, it can be directly inferred that the large critical curve of 
MACS J0717.5+3745 cannot be considered in tension with the $\Lambda$CDM model. This 
finding is supported by the results of the previous sections, which showed that the uncertainty of the 
mass function, particularly at the high-mass end, and the uncertainty in the shape of galaxy clusters 
allow a wide range of distributions. The recently reported higher redshift for the primary 
lensed system \citep{Limousin2011} and hence a smaller inferred Einstein radius would further strengthen 
our conclusions. Because MACS J0717.5+3745 is one of the most dynamically active known 
clusters and because \cite{Redlich2012} show that the distributions of the largest 
Einstein radius will be significantly shifted to larger values if dynamical mergers are accounted for, it 
can with certainty be deduced that the large Einstein radius of MACS J0717.5+3745 is consistent with the 
standard $\Lambda$CDM model.

\begin{figure}
\centering
\includegraphics[width=0.975\linewidth]{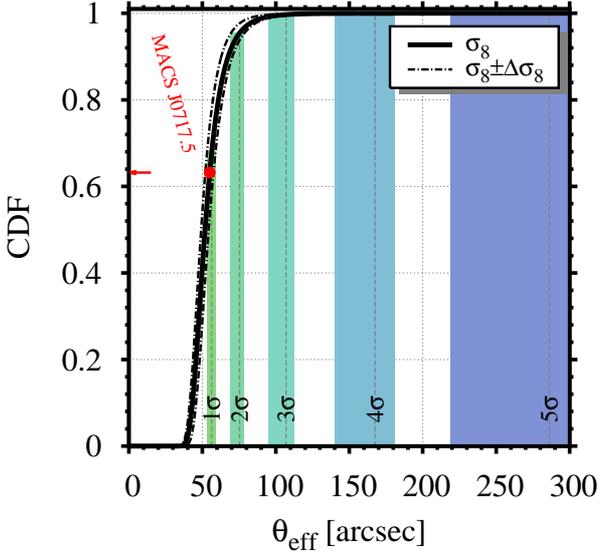}
\caption{CDF of the largest effective Einstein radius, assuming a redshift interval of $0.5\le z\le 1.0 $, the
full sky and a source redshift of $z_{\rm s}=2.5$. The dashed-dotted lines illustrate the uncertainty in the 
CDF due to the imprecision of the \textit{ WMAP7} value of $\sigma_8$. The vertical shaded regions denote 
the values of $\theta_{\rm eff}$ for which the CDF takes a value equal to $n\sigma$ with $n\in[1,5]$, as 
labelled in the panel. The red filled circle denotes the measured value of $\theta_{\rm eff}$ for MACS~J0717.5 
and the short red arrow points at the corresponding value of the CDF.}\label{fig:MACS_ER_confidence}
\end{figure}

With this in mind, these inferred occurrence probabilities should just be considered a rough estimate. 
Because of the uncertainties in modelling the distribution of Einstein radii, an 
observed critical curve should exhibit a much larger extent in order to be taken as clearly in tension with 
the $\Lambda$CDM model. To quantify this statement, we take the CDF of $\theta_{\rm eff}$ for 
MACS~J0717.5+3745 and calculate the values of $\theta_{\rm eff}$ for which the CDF takes values that 
correspond to confidence levels $n\sigma$ with $n\in[1,5]$, as shown in \autoref{fig:MACS_ER_confidence}. 
In order to lie beyond the $3\sigma$ level, corresponding to an occurrence probability of $\sim 0.3$ per 
cent, and to account for the uncertainty stemming from $\sigma_8$, $\theta_{\rm eff}$ should be larger 
than $\sim 115\,{\rm arcsec}$. This number will also be strongly affected by the redshift of the source 
population in the sense that sources at lower (higher) redshifts will result in a smaller (larger) value of 
$\theta^{3\sigma}_{\rm eff}$. Of course, the inclusion of dynamical mergers will increase this limit to 
even larger values.
\subsection{The probability of occurrence of the mass}
\begin{figure}
\centering
\includegraphics[width=0.975\linewidth]{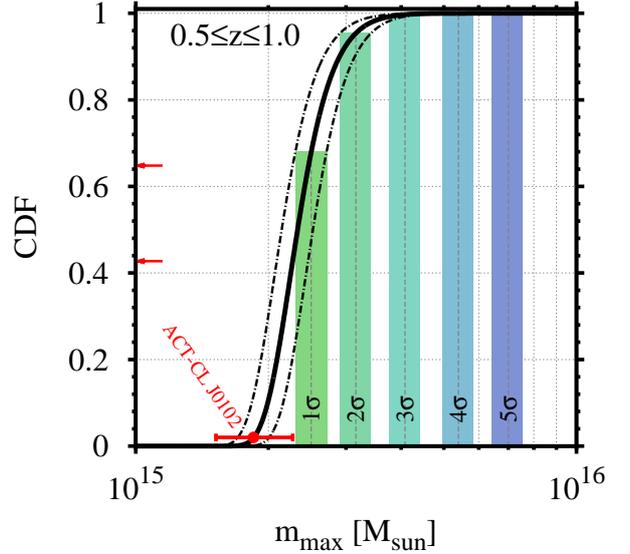}
\caption{CDF of the most massive halo that is expected to be found in the redshift interval of $0.5\le z\le 1.0 $ in 
an survey area of $A_{\rm s}=22\,735\,{\rm deg}^2$. The dashed-dotted lines illustrate the uncertainty in the CDF 
due to the imprecision of the \textit{ WMAP7} value of $\sigma_8$. The vertical shaded regions denote the values 
of the mass for which the CDF takes a value equal to $n\sigma$ with $n\in[1,5]$, as labelled in the panel. We also 
added for reference the mass of ACT-CL J0102 (red errorbar), the most massive known cluster in the given redshift range, 
where the two small red arrows point at the values of the CDF stemming from the upper allowed mass for the distributions 
based on $\sigma_8$ (lower arrow) and $\sigma_8-\Delta\sigma_8$ (upper arrow).}
\label{fig:MACS_cdf_mass}
\end{figure}

Apart from the larger critical curve, MACS J0717.5+3745 is also considered to be one of the most massive 
clusters in the redshift range of $0.5\le z\le1.0$. The strong-lensing analysis by \cite{Zitrin2009} revealed 
that the mass enclosed by the critical curve is $(7.4\pm 0.5)\times 10^{14}\,M_\odot$ and that the mass 
enclosed within the larger critical curve for the multiply-lensed dropout galaxy at $z\sim 4$ is found to be 
$\sim 1\times 10^{15}\,M_\odot$. These values are the masses in the innermost regions of the cluster, and 
thus the total mass can be considered to be a multiple of these values. Therefore, it is also interesting to study 
the expectation for the most massive galaxy cluster in the redshift range of interest of $0.5\le z\le 1.0 $.\\
EVS can also be applied to study the distribution of the most massive halo in a given volume 
\citep{Chongchitnan2012, Davis2011, Harrison&Coles2011, Harrison&Coles2012, Waizmann2011, Waizmann2012a, 
Waizmann2012b}; we will follow the procedure shown in \cite{Waizmann2012a} to compute the distribution 
function. The results are presented in \autoref{fig:MACS_cdf_mass}, where the CDF of the most massive halo in 
$0.5\le z\le 1.0 $ is shown for the nominal MACS survey area of $A_{\rm s}=22\,735\,
{\rm deg}^2$. For reference, we added ACT-CL~J0102-4915 \citep{Menanteau2011}, which is currently the most 
massive known cluster in the range of $0.5\le z\le 1.0$. The red filled circle with errorbars illustrates the mass 
and the upper/lower mass limits after the correction for the Eddington bias \citep{Eddington1913} according to 
\cite{Mortonson2011}. The two small red arrows point at the values of the CDF stemming from the upper allowed 
mass for the distributions based on $\sigma_8$ (lower arrow) and $\sigma_8-\Delta\sigma_8$ (upper arrow). 
As already discussed in \cite{Waizmann2012a}, this cluster can be considered to be in agreement with $\Lambda$CDM. 
The vertical, blue shaded areas indicate the values of the maximum halo mass for which the CDF corresponds to 
the values equal to the labelled confidence level. In order to let MACS J0717.5+3745 be in tension with 
$\Lambda$CDM, its mass should lie at least above the value indicated by the $3\sigma$ area, which would 
correspond to an occurrence probability of $\lesssim 0.3$ per cent.

Therefore, the mass that should be at least exceeded is $M_{3\sigma}\simeq 4\times 10^{15}\,M_\odot$, which is 
\textit{per se} already a very high mass for a cluster and, since cluster masses are subject to significant uncertainties, 
the lower mass limit should lie above this value. Furthermore, for a statistical analysis similar to the one of 
ACT-CL~J0102-4915, one also has to shift the observed mass to a smaller value to account for the Eddington bias 
\citep{Mortonson2011}. 

Considering the complex dynamical state of MACS~J0717.5+3745, the embedding in a large scale filament and the 
very high mass that would be required, it seems to be more than doubtful that, despite the large mass enclosed in 
the critical curve \citep{Zitrin2009},  the total mass of this system can be used to exclude $\Lambda$CDM. When 
summing up our findings for both the Einstein radius and the mass, we find that the characteristics of 
MACS~J0717.5+3745 are not unlikely to be found in a $\Lambda$CDM framework. These results are substantially 
different from the findings of \cite{Zitrin2009}, who report that the probability to find such a system is of the 
order of $\sim 10^{-7}$. The main reason for this difference is that in order to sample the distribution of the maxima, 
a larger number of universes ($\sim 1000$) have to be simulated, which is feasible with a semi-analytic approach but 
where a N-body based approach \citep{Broadhurst&Barkana2008} falls short. The latter can be used to infer the statistical 
characteristic of large Einstein radii in general but not of the single largest observation. Furthermore, one biases the results 
by a posteriori choosing the redshift interval and the assigned Einstein radius, since it can not be known before at which 
redshift the most massive cluster will be realised and what Einstein radius it will have \citep[see e.g.][]{Hotchkiss2011, 
Waizmann2012a}. 
\section{Summary and conclusions}\label{sec:conclusions}
In this work, we have presented a study of the distribution of the single largest Einstein radius at redshifts 
$0.5\le z\le 1.0$, based on the MC simulation of triaxial halo populations in mock universes extending the work of 
OB09. The details of the implementation of our semi-analytic method can be found in \cite{Redlich2012}. Our work 
can be divided into three distinct parts: first, a preparatory study; second, a study of the effects that impact on the 
distribution of the maximum Einstein radius; and third, a case study for MACS~J0717.5+3745.

In the first preparatory part, we showed that $\sim 1000$ mock universes are sufficient to sample the distribution of the 
maxima and that the resulting distribution can be very well fitted by the functional shape of the generalized extreme 
value distribution. In general, we find that the distribution of the largest Einstein radii can be well described by a 
Fr\'{e}chet (Type II) distribution, indicating that the distribution is bound from below. Furthermore, we confirm the 
findings of OB09 that the sample of maxima is distributed in a wide range in the mass--redshift plane, indicating 
that the single largest Einstein radius has its origin by no means necessarily in the most massive haloes. This indicates 
that the triaxiality, together with the halo orientation with respect to the observer, has a stronger impact than the mass 
of the cluster itself. We also report that different definitions of the Einstein radius, like the effective or the median 
one, can lead to the selection of different haloes for the largest Einstein radius, particularly for smaller values of the 
maximum Einstein radius. However, both definitions lead to identical results for the largest realizations. 

In the second part, we studied the influence of different underlying assumptions and effects on the resulting extreme 
value distributions. The results of this part can be summarised as follows.
\begin{itemize}
\item \textit{Mass function.} We sampled the extreme value distribution for four different mass functions, comprising 
the \cite{Press1974}, \cite{Sheth&Tormen1999}, \cite{Tinker2008} and \cite{Crocce2010} mass functions. Modifying the 
size of the halo sample in each mock universe from which the maximum Einstein radius will be sampled, the different 
mass functions lead mainly to a shift to smaller or larger Einstein radii, while the impact on the shape of the distribution 
is less pronounced. Of course, the effect of a different normalisation of the matter power spectrum, $\sigma_8$, is 
similar in nature.
\item \textit{Triaxiality.} We studied the impact of triaxiality by introducing different cut-offs in the underlying axis-ratio 
distribution, with the result that the extreme value distribution of the largest Einstein radius is very sensitive to the presence of 
small axis ratios and hence, very elongated objects. The different cut-offs lead not only to shifts of the resulting extreme 
value distributions, but also substantially influence the shape of the distribution. The more elongated objects 
are allowed to exist, the higher will be the tail of the extreme value distributions towards very large values of the Einstein radius.
\item \textit{Inner slope and $c$--$M$ relation.} Both underlying assumptions impact on the resulting distributions and 
exhibit a particular behaviour, which is caused by projection effects. Since the extreme value distribution will be naturally 
based on elongated haloes that are oriented along the line of sight, the strong-lensing signal can be significantly enhanced 
from the outer regions of the halo. It remains unclear if and where a radius cut-off of the density profile, e.g. at the virial 
radius \citep[see e.g.][]{Oguri&Keeton2004, Baltz2009}, should be imposed. This adds an additional uncertainty to the results 
that have been listed so far.
\end{itemize}
With our study, we could show that a multitude of underlying assumptions strongly influence the extreme value distribution of 
the largest Einstein radius. Many of those require more detailed studies, e.g. the triaxiality of dark matter haloes. Another 
effect having a strong impact on the extreme value distribution is the presence of dynamical mergers as shown in the work 
of \cite{Redlich2012}, and it will be studied by the authors in further detail in future work. In view of this complexity, it is 
unlikely that the extreme value distribution of the Einstein radius can be used for consistency test of $\Lambda$CDM. 
However, due to its enhanced sensitivity to the underlying assumptions, it could very well be used to learn more about these 
assumptions.

In the last part of this work, we used the previously studied extreme value distributions to assess the probability of occurrence 
for the largest known Einstein radius of MACS~J0717.5+3745 \citep{Zitrin2009}. Accounting only for the uncertainty in $\sigma_8$, 
we find for the observed effective Einstein radius of $\theta_{\rm eff}=55\pm3\,{\rm arcsec}$ an occurrence probability of 
$\sim(11-42)$ per cent for the MACS survey area and of $\sim(18-61)$ per cent on the full sky, indicating that this observation 
can not be considered in conflict with $\Lambda$CDM. This conclusion is supported by the fact that the probability range would 
widen further if we would account for the uncertainties in the underlying assumptions, for instance, mass function and triaxiality, 
rendering any claim of tension with $\Lambda$CDMuntenable. Furthermore, we neglected the impact of dynamical merging for which 
MACS~J0717.5+3745 is a prime example, which again would make extremely large critical curves more likely to be found.

However, apart from our results for the large Einstein radius, MACS~J0717.5+3745 is a candidate for the most massive known 
galaxy cluster in the redshift range of $0.5\le z\le 1.0$, as indicated by the mass enclosed by the critical curve of $\sim 1\times 
10^{15}\,M_\odot$ \citep{Zitrin2009}. Since this is the mass that is contained only in the innermost region, the overall cluster mass is expected to 
be significantly larger\footnote{It should be mentioned that the recent strong lensing analysis of MACS~J0717.5+3745 by 
\cite{Limousin2011} finds a redshift for the primary lensed galaxy of $z_{\rm s}\sim 2.96$ instead of $z_{\rm s}\sim 2.5$ used 
by \cite{Zitrin2009}. This change would lead to a shift in the surface mass normalization, $\kappa$, and hence to a smaller mass 
estimate.}. A more thorough mass estimate for  MACS~J0717.5+3745 is expected to be provided by the Cluster Lensing And Supernova 
survey with Hubble (CLASH) \citep{Postman2012}. Inspired by this result, we calculated the total mass a galaxy cluster would need to 
have  in the redshift range of $0.5\le z\le 1.0$ in order to exhibit significant tension with $\Lambda$CDM. To do so, we utilised the 
extreme value statistical approach for halo masses used in \cite{Waizmann2011} and found that for a $3\sigma$ deviation from the 
$\Lambda$CDM model, the cluster would need to have a mass of at least $M_{3\sigma}\simeq 4\times 10^{15}\,M_\odot$. This 
value needs to be even larger in order to account for the correction for the Eddington bias \citep[see e.g.][]{Mortonson2011} and the 
unavoidable uncertainties in the mass determination. Whether the mass for MACS~J0717.5+3745 will reach such high values remains 
to be seen.

As a closing remark, we conclude that it seems to be more than doubtful that the single largest observed Einstein radius can be used 
as a basis for $\Lambda$CDM falsification experiments. However, we expect nevertheless useful insights into the underlying assumptions 
that enter the modelling of the Einstein radius distribution. In the future, we intend to perform further studies along these lines. 
\begin{acknowledgements}
We would like to thank Adi Zitrin and Marceau Limousin for the very helpful discussions. 
JCW acknowledges financial contributions from the contracts ASI-INAF I/023/05/0, 
ASI-INAF I/088/06/0, ASI I/016/07/0 COFIS, ASI Euclid-DUNE I/064/08/0, ASI-Uni 
Bologna-Astronomy Dept. Euclid-NIS I/039/10/0, and PRIN MIUR 2008 \textit{Dark 
energy and cosmology with large galaxy surveys}. MR thanks the Sydney Institute for 
Astronomy for the hospitality and the German Academic Exchange Service (DAAD) for 
the financial support. Furthermore, MR's work was supported in part by contract 
research \textit{Internationale Spitzenforschung II-1} of the Baden-W\"{u}rttemberg 
Stiftung. MB is supported in part by the Transregio-Sonderforschungsbereich \textit{The Dark 
Universe} of the German Science Foundation.
\end{acknowledgements}

\bibliographystyle{aa}
\bibliography{GEV_SL}

\end{document}